\documentclass[aip,rsi,reprint]{revtex4-1}

\newcommand{\mrowtabcel}[2]{\begin{tabular}{@{}#1@{}}#2\end{tabular}}
\usepackage{times}
\usepackage{subfig}
\usepackage{graphicx}
\usepackage{epstopdf}
\usepackage{multirow}
\usepackage{paralist}
\usepackage[tbtags]{amsmath}
\usepackage{amssymb}
\usepackage{amsthm} 
\usepackage{bm}
\usepackage{times}
\usepackage{mathptmx}
\usepackage{siunitx}
\usepackage{upgreek}
\usepackage{tabularx}
\usepackage{indentfirst}
\newtheoremstyle{query}%
{}{}
{\color{red}}
{}
{\sffamily\bfseries}{:}{12pt}
{}
\theoremstyle{query}

\allowdisplaybreaks[4]
\begin{document}
\preprint{AIP/123-QED}

\title{Mathematic Model And Error Analysis of Moving-base Rotating Accelerometer Gravity Gradiometer}
\author{MINGBIAO YU}
\email[]{ymb\_moon@126.com}
\noaffiliation
\author{TIJING CAI}
\email[Corresponding author:]{caitij@seu.edu.cn}
\noaffiliation
\affiliation{Instrument Science and Engineering College, Southeast University, Nanjing 210096, China}

\date{\today}

\begin{abstract}
 In moving-base gravity gradiometry, accelerometer mounting errors and mismatch cause a rotating accelerometer gravity gradiometer (RAGG) to be susceptible to its own motion. In this study, we comprehensively consider accelerometer mounting errors, circuit gain mismatch, accelerometer linear scale factors imbalances, accelerometer second-order error coefficients and construct three RAGG models, namely a numerical model, an analytical model, and a simplified analytical model. The analytical model and the simplified analytical model are used to interpret the error propagation mechanism and develop error compensation techniques.
A multifrequency gravitational gradient simulation experiment and a dynamic simulation experiment are designed to verify the correctness of the
three RAGG models; three turbulence simulation experiments are designed to evaluate the noise floor of the analytical models at different intensity of air turbulence. The mean of air turbulence is in the range of 70 $\sim$ 230~mg, the noise density of the analytical model
 is about 0.13 Eo/$\surd$Hz, and that of the simplified analytical model is in the range of 0.25 $\sim$ 1.24 Eo/$\surd$Hz.
 The noise density of the analytical models is far less than 7~Eo/$\surd$Hz, which suggests that using the error compensation techniques based on the analytical models, the turbulence threshold of survey flying may be widened from current 100~mg to 200~mg.
\end{abstract}
\maketitle

\section{Introduction}
Airborne gravity gradiometry is an advanced technology for surveying a gravity field; it acquires gravity field information with high efficiency and high spatial resolution. Compared with  gravity information, the gravity gradient tensor provides more  information on the field source such as orientation, depth, and shape~\cite{Tang2018,Yan2015}. The world's first airborne gravity gradiometry was performed using the Falcon-AGG system in October 1999. Airborne gravity gradiometry has now been conducted for nearly 20 years, and the experience gained in airborne gravity gradiometry, and the analysis and interpretation of gravity gradient data have greatly promoted  developments in geological science, resource exploration, high-precision navigation, and related fields~\cite{Kahn1985IToGaRS,welker2013gravity,araya2012gravity,Jekeli2006Precision,rogers2009investigation,annecchione2007benefits,66158}.
The application value of gravity gradiometry has been recognized, and the associated technology and data interpretation have become  of interest in scientific research, which has promoted  development of gravity gradiometry.
 There are many different gravity gradiometers  under development, for example: rotating accelerometer gravity gradiometers, superconducting gravity gradiometers, cold atomic interferometer gravity gradiometers,
  MEMS gravity gradiometers, and gravitec gravity gradiometers\cite{Paik2007Geodesy,anstie2009vk,difrancesco2007advances,hao2013design,moody1983preliminary,moody2011superconducting,liu2014design}. However, rotating accelerometer gravity gradiometers are the only type successfully used in airborne surveys; all other types are either in fight testing or in a laboratory setting\cite{rogers2009investigation}. Companies that operate commercial rotating accelerometer gravity gradiometer systems are: Bell Geospace (3D-FTG), ARKeX (FTGeX), GEDEX (HD-AGG), and FUGRO (AGG-Falcon).

A rotating accelerometer gravity gradiometer (RAGG) was developed by Ernest Metzger of Bell-Aerospace in the 1980s. The minimum configuration of a RAGG consists of two pairs of high-quality, low-noise, matched accelerometers, which are equi-spaced on a rotating disc with their sensitive axes tangential to the disc. The spin axis of the disc is perpendicular to the plane of the disc, and passes through its center\cite{lee2001falcon,HaibingLi2010}. The RAGG measures the gradients in the disc plane (RAGG input plane). The disc rotates at a constant speed, typically 0.25 Hz; this rotation results in the gravity gradient signal being modulated at 0.5 Hz, while the linear accelerations in the disc plane are modulated at 0.25 Hz. If the accelerometers are perfectly mounted and scale-factor balanced and the second-order error coefficients are small enough, the sum of the diametrically opposed accelerometer reject linear accelerations in the disc plane, and the difference of the sum of the two pairs accelerometer cancel out the angular acceleration about the spin axis and zero bias \cite{yu2018calibration}. Since a small misalignment error of $10^{-4}$ rad will make the RAGG sensitive to the linear accelerations, as the material ages, the influence of accelerometer mounting errors and imbalances in accelerometer scale factors cannot be ignored. Techniques such as automatic on-line continuous scale-factor imbalance, second-order error coefficient compensation, and mounting error compensation are required\cite{metzger1977recent,heard1988gravity}. Meanwhile, high rate post mission compensation further compensates the measurement error caused by the motion of the RAGG\cite{geospace2004final}.
 To ensure the gravity gradient information to be used in desirable applications, the noise level of 7 Eo/$\surd$Hz in the dynamic environment of survey flying is desirable. Good weather conditions should generally be chosen in airborne gravity gradiometry to limit turbulence experienced during flight.
 Dransfield reports the effect of turbulence on current gravity gradiometer: noise levels of an FTG GGI is about 13 $\sim$ 23 Eo/$\surd$Hz at 12 $\sim$ 40~mg, and that of a FALCON GGI is about 3 $\sim$ 4 Eo/$\surd$Hz at 28 $\sim$ 64~mg\cite{dransfield2013performance}. Currently, the turbulence threshold of airborne gravity gradiometry is about 70 $\sim$ 100~mg.

Jekeli \cite{jekeli2006airborne} analyzed the requirements of gyro precision for moving base gravity gradiometer with different sensitivities.
Ma \cite{Ma2012}  analyzed the error terms with a one-factor-at-a-time method, but did not consider the coupling error of each error term synthetically and did not obtain a RAGG output model for all  sources of errors.
 Here, we synthetically consider circuit gain mismatch, installation errors, accelerometer scale-factor imbalance, and accelerometer second-order error coefficients, and deduce three RAGG models: a numerical model, a analytic model, and a simplified analytic model. From the analytical models, we can obtain  error propagation coefficients for the motion of the RAGG, and determine the relationships among error propagation coefficients, installation errors, scale-factor imbalance, circuit gain mismatch, etc. The analytical models can interpret the error propagation mechanism of the RAGG and help to develop error compensation techniques. The RAGG numerical model is a virtual RAGG with a comprehensive set of precisely adjustable parameters; based on it, many key techniques of the RAGG, such as automatic online continuous error compensation, post error compensation, and self-gradient modeling, can be verified.

\section{Models of the RAGG}
\subsection{RAGG Analytical Model}
\subsubsection{Accelerometer installation error and output model}\label{sec2.1}
\begin{figure}[!t]\centering
	\includegraphics[width=0.4\textwidth]{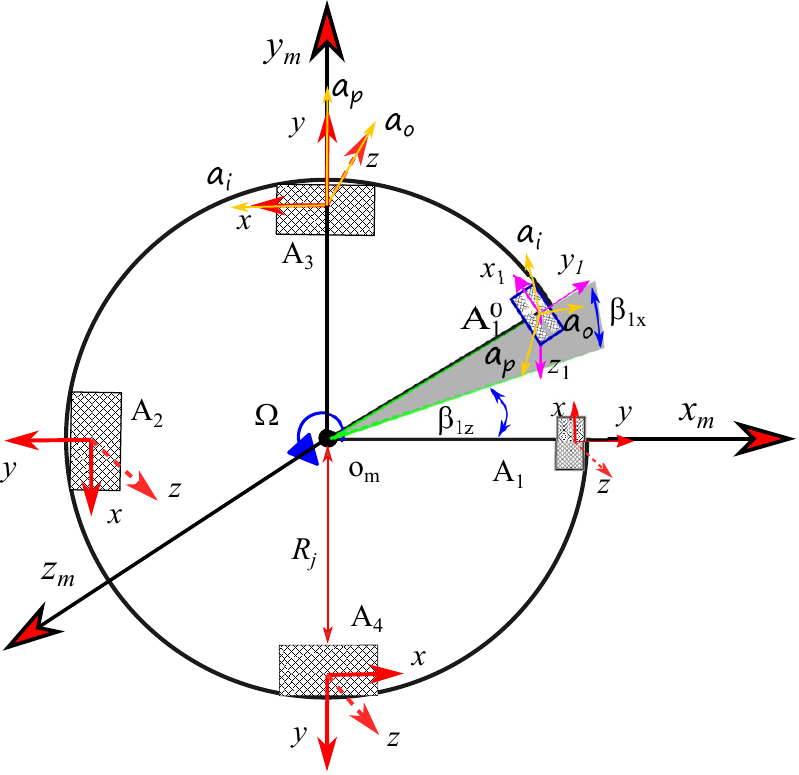}
	\caption{RAGG accelerometer installation errors.}\label{fig1}
\end{figure}

To simplify description of the installation error, we first define the RAGG measurement frame. In Fig. \ref{fig1}, the origin of the RAGG measurement frame (${{o}_{m}}$) is at the center point of the disc, and its x- and y-axes point respectively to the initial positions A1 and A3 of the accelerometer; its z-axis coincides with the spin axis of the disc. The RAGG measurement frame is a space-fixed coordinate system, and does not rotate with the rotating disc.

The accelerometer mounting errors consist of mounting position errors and input-axis misalignments.
For the sake of clarity, we take the accelerometer ${A_1}$ as an example for the mounting errors.
In Fig.\ref{fig1}, $A_1$, $A_2$, $A_3$ and $A_4$ represent the nominal mounting positions,
$A_{1}^{0}$ represents the actual mounting position of the accelerometer $A_1$ and the deviation from the nominal installation point. Another three reference coordinate systems are adopted: the accelerometer nominal frame of the nominal mounting position ($xyz$, axes marked in red), the accelerometer nominal frame of the actual mounting position (${{x}_{1}}{{y}_{1}}{{z}_{1}}$, axes marked in magenta),  and the accelerometer measurement frame (${{a}_{i}}{{a}_{o}}{{a}_{p}}$, axes marked in orange). {The accelerometer nominal frame of the nominal mounting position ($xyz$, axes marked in red) and the accelerometer nominal frame of the actual mounting position (${{x}_{1}}{{y}_{1}}{{z}_{1}}$, axes marked in magenta) are all the accelerometer nominal frame, these two reference coordinate systems are named after the location of the origin: the accelerometer nominal mounting position and the accelerometer actual mounting position.}
The origin of the accelerometer nominal frame is located at the accelerometer mounting position; its x-axis is tangential to the disc along the rotating direction, and its y-axis is from the disc center to the accelerometer position along the radial direction. Among the four frames, only the accelerometer measurement frame (${{a}_{i}}{{a}_{o}}{{a}_{p}}$) and the accelerometer nominal frame of the actual accelerometer mounting position (${{x}_{1}}{{y}_{1}}{{z}_{1}}$) are concentric frames. The accelerometer mounting position error is the position difference between the actual mounting position and the nominal mounting position. Misalignment error is the orientation deviation between the accelerometer input axis (${{a}_{i}}$) and the tangential direction of the actual accelerometer  mounting position (${{x}_{1}}$).
We can use three parameters to determine the accelerometer mounting position with respect to the nominal mounting position: radial distance, initial phase angle, and altitude angle. The radial distance is the distance from the disc center to the accelerometer mounting position. We use the notation ${{R}_{j}}$ to denote the radial distance of accelerometer ${{A}_{j}}$. The radial distance ${{R}_{j}}$ of accelerometer ${{A}_{j}}$ can also be expressed as ${{R}_{j}}=R+d{{R}_{j}}$, where $d{{R}_{j}}$ is the radial distance error of accelerometer ${{A}_{j}}$, R is accelerometer nominal mounting distance. The nominal mounting positions of the four accelerometers are in the same plane, and we define this as the reference plane. The central angle from the accelerometer nominal mounting position to the projection of the actual accelerometer mounting position on the reference plane is defined as the initial phase angle. The notation ${{\beta }_{jz}}$ denotes the initial phase angle of accelerometer ${{A}_{j}}$. If the direction vector from the accelerometer nominal mounting position to the actual mounting position coincides with the rotating direction of the disc, then the initial phase angle ${{\beta }_{jz}}$ is positive; otherwise, the angle is negative. The angle between the radial distance line and the reference plane is defined as the altitude angle. The notation ${{\beta }_{jx}}$ represents the altitude angle of  accelerometer ${{A}_{j}}$. If the z-coordinate of the actual accelerometer  mounting position in the RAGG measurement frame is positive, then its corresponding altitude angle is positive; otherwise, the angle is negative.

The second type of mounting error is a misalignment error due to the orientation deviation between the accelerometer sensitive axis (${{a}_{i}}$) and the tangential direction of the disc (${{x}_{1}}$). If we rotate the accelerometer measurement frame about its y-axis by $-{{\vartheta }_{jy}}$ and then about its z-axis by $-{{\vartheta }_{jz}}$, then the input axis (${{a}_{i}}$) will coincide with the x-axis of the accelerometer nominal frame of the actual mounting position (tangential direction of the disc). So we use these two angles ${{\vartheta }_{jy}}$ and ${{\vartheta }_{jz}}$ as the misalignment error parameters. From the above, we can use the five parameters (${{R}_{j}}$, ${{\beta }_{jx}}$, ${{\beta }_{jz}}$, ${{\vartheta }_{jy}}$, and ${{\vartheta }_{jz}}$) to determine the accelerometer mounting error of the RAGG.

 The accelerometer output is the response of its applied specific force:
\begin{equation}\label{eq1}
\begin{gathered}
  {{{I_j}} \mathord{\left/
 {\vphantom {{{I_j}} {{K_{j1}}}}} \right.
 \kern-\nulldelimiterspace} {{K_{j1}}}} = {f_{ji}} + {K_{j0}} + {K_{j2}}f_{ji}^2 + {K_{j5}}f_{jo}^2 + {K_{j7}}f_{jp}^2
 \\
   + {K_{j4}}{f_{ji}}{f_{jP}} + {K_{j6}}{f_{ji}}{f_{jo}} + {K_{j8}}{f_{jo}}{f_{jp}}\thickspace.
\end{gathered}
\end{equation}
Where $I_j$ is the electrical current output of  accelerometer $A_j$; ${{f}_{ji}}$, ${{f}_{jo}}$, ${{f}_{jp}}$ are the applied specific forces in the directions of the input, output, and pendulous axes, respectively; ${{K}_{j1}}$ is the linear scale factor (in units of A/g); ${{K}_{j0}}$ is the null bias (in units of g); and ${{K}_{j2}}$, ${{K}_{j4}}$, ${{K}_{j5}}$, ${{K}_{j6}}$, ${{K}_{j7}}$, ${{K}_{j8}}$ are the second-order error coefficients (in units of g/g$^2$).
We introduce another angle ${{\vartheta }_{jx}}$, using three small angles ${{\vartheta }_{jx}}$, ${{\vartheta }_{jy}}$, and ${{\vartheta }_{jz}}$, to describe the misalignment between the accelerometer measurement frame and the accelerometer nominal frame of the actual mounting position. The accelerometer measurement frame results from the rotation of the accelerometer nominal frame of the actual mounting position first about the x-axis by ${{\vartheta }_{jx}}$, second about the y-axis by ${{\vartheta }_{jy}}$, and then
about the z-axis by ${{\vartheta }_{jz}}$.
The transformation matrix from the accelerometer nominal frame of the actual mounting position to the accelerometer measurement frame is:
\begin{equation}\label{eq2}
\bm{C}=\left[ \begin{matrix}
   1 & {{\vartheta }_{jz}} & -{{\vartheta }_{jy}}  \\
   -{{\vartheta }_{jz}} & 1 & {{\vartheta }_{jx}}  \\
   {{\vartheta }_{jy}} & -{{\vartheta }_{jx}} & 1  \\
\end{matrix} \right].
\end{equation}
Let ${{f}_{jx}}$, ${{f}_{jy}}$, ${{f}_{jz}}$  denote the coordinates of the specific force of  accelerometer ${{A}_{j}}$ in the accelerometer nominal frame of the actual mounting position. So, we have:
\begin{equation}\label{eq3}	
\begin{array}{l}
{\rm{ }}{f_{ji}} = {f_{jx}} - {\rm{ }}{f_{jz}}{\vartheta _{jy}}{\rm{ }} + {\rm{ }}{f_{jy}}{\vartheta _{jz}}\\
{\rm{ }}{f_{jo}} = {f_{jy}}{\rm{ }} + {\rm{ }}{f_{jz}}{\vartheta _{jx}} - {\rm{ }}{f_{jx}}{\vartheta _{jz}}\\
{\rm{ }}{f_{jp}} = {f_{jz}}{\rm{ }} - {\rm{ }}{f_{jy}}{\vartheta _{jx}}{\rm{ }} + {\rm{ }}{f_{jx}}{\vartheta _{jy}} \thickspace .
\end{array}
\end{equation}	
Substituting  Eq.\eqref{eq3} into Eq.\eqref{eq1}, we get:
 \begin{equation}\label{eq2-5a}
\begin{array}{l}
{{{I_j}} \mathord{\left/
 {\vphantom {{{I_j}} {{K_{j1}}}}} \right.
 \kern-\nulldelimiterspace} {{K_{j1}}}} = {f_{jx}} + {f_{jy}}\vartheta _{jz}^* - {f_{jz}}\vartheta _{jy}^* + {K_{j0}} + K_{j2}^*{f_{jx}}^2 + K_{j5}^*{f_{jy}}^2\\
 + K_{j7}^*{f_{jz}}^2 + K_{j6}^*{f_{jx}}{f_{jy}} + K_{j4}^*{f_{jx}}{f_{jz}} + K_{j8}^*{f_{jy}}{f_{jz}}
\end{array}
 \end{equation}
Where $\vartheta _{jz}^*$, $\vartheta _{jy}^*$, $K _{j2}^*$, $K _{j4}^*$, $K _{j5}^*$, $K _{j6}^*$, $K _{j7}^*$, $K _{j8}^*$ are given :
\begin{align}\label{eq2-5b}
\vartheta _{jz}^* &= ({\vartheta _{jz}} + {\vartheta _{jx}}{\vartheta _{jy}}) \notag\\
\vartheta _{jy}^* &= {\vartheta _{jy}} - {\vartheta _{jx}}{\vartheta _{jz}}\notag\\
K_{j2}^* &= {K_{j2}} + {K_{j4}}{\vartheta _{jy}} - {K_{j6}}{\vartheta _{jz}} + {K_{j7}}\vartheta _{jy}^2 - {K_{j8}}{\vartheta _{jy}}{\vartheta _{jz}}
+ {K_{j5}}\vartheta _{jz}^2\notag\\
K_{j4}^* &= (1 - \vartheta _{jy}^2 + {\vartheta _{jx}}{\vartheta _{jy}}{\vartheta _{jz}}){K_{j4}} + 2({\vartheta _{jx}}{\vartheta _{jz}} - {\vartheta _{jy}}){K_{j2}}\notag\\&
 - 2({\vartheta _{jy}}\vartheta _{jz}^2 + {\vartheta _{jx}}{\vartheta _{jz}}){K_{j5}} + ({\vartheta _{jx}} + 2{\vartheta _{jy}}{\vartheta _{jz}} \notag\\&- {\vartheta _{jx}}\vartheta _{jz}^2){K_{j6}}
 + 2{\vartheta _{jy}}{K_{j7}} + (\vartheta _{jy}^2{\vartheta _{jz}} + {\vartheta _{jx}}{\vartheta _{jy}} - {\vartheta _{jz}}){K_{j8}}\notag\\
K_{j5}^* &= {K_{j5}}(1 - 2{\vartheta _{jx}}{\vartheta _{jy}}{\vartheta _{jz}} + \vartheta _{jx}^2\vartheta _{jy}^2\vartheta _{jz}^2) - {K_{j8}}({\vartheta _{jx}} - \vartheta _{jx}^2{\vartheta _{jy}}{\vartheta _{jz}}) \notag\\&
+ {K_{j7}}\vartheta _{jx}^2 - {K_{j4}}({\vartheta _{jx}}{\vartheta _{jz}} + \vartheta _{jx}^2{\vartheta _{jy}}) + {K_{j2}}(2{\vartheta _{jx}}{\vartheta _{jy}}{\vartheta _{jz}} + \vartheta _{jz}^2 \notag\\&
 + \vartheta _{jx}^2\vartheta _{jy}^2)
 + {K_{j6}}({\vartheta _{jz}} + {\vartheta _{jx}}{\vartheta _{jy}} - {\vartheta _{jx}}{\vartheta _{jy}}\vartheta _{jz}^2 - \vartheta _{jx}^2\vartheta _{jy}^2{\vartheta _{jz}})\notag\\
K_{j6}^* &= (1 - \vartheta _{jz}^2 - 2{\vartheta _{jx}}{\vartheta _{jy}}{\vartheta _{jz}}){K_{j6}} + (2{\vartheta _{jz}} + 2{\vartheta _{jx}}{\vartheta _{jy}}){K_{j2}}\notag\\&
 + ({\vartheta _{jx}}\vartheta _{jy}^2 + {\vartheta _{jy}}{\vartheta _{jz}} - {\vartheta _{jx}}){K_{j4}} - 2{\vartheta _x}{\vartheta _y}{K_7}\notag\\&
 + (2{\vartheta _x}{\vartheta _y}\vartheta _z^2 - 2{\vartheta _z}){K_5} + ({\vartheta _y} + {\vartheta _x}{\vartheta _z} - {\vartheta _x}\vartheta _y^2{\vartheta _z}){K_8}\notag\\
K_{j7}^* &= {K_{j7}} + (\vartheta _{jx}^2\vartheta _{jz}^2 - 2{\vartheta _{jx}}{\vartheta _{jy}}{\vartheta _{jz}} + \vartheta _{jy}^2){K_{j2}} + ({\vartheta _{jx}}{\vartheta _{jz}} \notag\\& - {\vartheta _{jy}}){K_{j4}}
 + (\vartheta _{jx}^2{\vartheta _{jz}} + {\vartheta _{jx}}{\vartheta _{jy}}\vartheta _{jz}^2 - {\vartheta _{jx}}{\vartheta _{jy}} - \vartheta _{jy}^2{\vartheta _{jz}}){K_{j6}}\notag\\&
 + (\vartheta _{jx}^2 + 2{\vartheta _{jx}}{\vartheta _{jy}}{\vartheta _{jz}} + \vartheta _{jy}^2\vartheta _{jz}^2){K_{j5}} + ({\vartheta _{jx}} + {\vartheta _{jy}}{\vartheta _{jz}}){K_{j8}}\notag\\
K_{j8}^* &= (1 - \vartheta _{jx}^2 - 2{\vartheta _{jx}}{\vartheta _{jy}}{\vartheta _{jz}}){K_{j8}}\notag\\&
 + 2({\vartheta _{jx}}\vartheta _{jz}^2 - {\vartheta _{jx}}\vartheta _{jy}^2 - {\vartheta _{jy}}{\vartheta _{jz}} + \vartheta _{jx}^2{\vartheta _{jy}}{\vartheta _{jz}}){K_{j2}}\notag\\&
 + ({\vartheta _{jz}} + 2{\vartheta _{jx}}{\vartheta _{jy}} - {\vartheta _{jz}}\vartheta _{jx}^2){K_4} - 2{\vartheta _{jx}}{K_{j7}}\notag\\&
 + (2{\vartheta _{jx}} + 2{\vartheta _{jy}}{\vartheta _{jz}} - 2\vartheta _{jx}^2{\vartheta _{jy}}{\vartheta _{jz}} - 2{\vartheta _{jx}}\vartheta _{jy}^2\vartheta _{jz}^2){K_{j5}}\notag\\&
 + (\vartheta _{jx}^2{\vartheta _{jy}} - \vartheta _{jx}^2{\vartheta _{jy}}\vartheta _{jz}^2 + 2{\vartheta _{jx}}{\vartheta _{jz}}\vartheta _{jy}^2 + 2{\vartheta _{jx}}{\vartheta _{jz}} \notag\\& + {\vartheta _{jy}}\vartheta _{jz}^2 - {\vartheta _{jy}}){K_{j6}}
\end{align}

The second-order error coefficients ${{K}_{j2}}$, ${{K}_{j4}}$, ${{K}_{j5}}$, ${{K}_{j6}}$, ${{K}_{j7}}$, ${{K}_{j8}}$ are of the order of ${{10}^{-6}}$ g/g$^2$ and the misalignment angles ${{\vartheta }_{jz}}$, ${{\vartheta }_{jy}}$, ${{\vartheta }_{jz}}$ are of the order of ${{10}^{-4}}$ rad;
thus, ${{K}_{ji}}{{\vartheta }_{jx}},i=2,4,5,6,7,8$, ${{K}_{ji}}{{\vartheta }_{jy}}$, and ${{K}_{ji}}{{\vartheta }_{jz}}$ are of the order of ${{10}^{-10}}$ g/g$^2$.
That is, the specific force is of magnitude 0.1 g, terms such as $K_{ji}\vartheta _{jy}f_{jx}f_{jy}$, $K_{ji}\vartheta _{jx}f_{jx}f_{jz}$ are of the order of  ${{10}^{-12}}$ g; therefore,
as for a gravity gradiometer with resolution 1 Eo, these terms can be neglected. Eq.\eqref{eq2-5a} becomes:
\begin{equation}\label{eq4}
\begin{gathered}
  {{{I_j}} \mathord{\left/
 {\vphantom {{{I_j}} {{K_{j1}}}}} \right.
 \kern-\nulldelimiterspace} {{K_{j1}}}} = {f_{jx}} + {f_{jy}}{\vartheta _{jz}} - {f_{jz}}{\vartheta _{jy}} + {K_{j0}} + {K_{j2}}{f_{jx}}^2 + {K_{j5}}{f_{jy}}^2 \\
   + {K_{j7}}{f_{jz}}^2 + {K_{j6}}{f_{jx}}{f_{jy}} + {K_{j4}}{f_{jx}}{f_{jz}} + {K_{j8}}{f_{jy}}{f_{jz}} \thickspace.\\
\end{gathered}
\end{equation}
The Eq.\eqref{eq4} is a approximation of the Eq.\eqref{eq2-5a}. Because the Eq.\eqref{eq4} has the some form with Eq.\eqref{eq2-5a}, no matter which equation is used, the derived RAGG analytical model will have the same form. To simplify the description, we use the Eq.\eqref{eq4} as the accelerometer output model to derive the RAGG output model.

The accelerometer mounted on the moving base RAGG is a flexible force rebalancing accelerometer. There are two types of force rebalancing accelerometer: a voltage one and a current one. The voltage-type force rebalancing accelerometer uses electrostatic actuation, and its output is a voltage signal. The current-type force rebalancing accelerometer uses electromagnetic actuation, and its output is a current signal. Electrostatic actuation inherently creates less heat and generates less thermal drift than electromagnetic actuation, but electrostatic actuation has a very small displacement range. Electrostatic actuation accelerometers (voltage-type accelerometers) with small measurement ranges are mainly used in spaceborne gravity gradiometers. Electromagnetic actuation accelerometers (current-type accelerometers) are usually applied to  airborne gravity gradiometers. As the output current of the RAGG accelerometer needs to be converted into a voltage signal for further processing,  we combine the current to voltage gain into the accelerometer model:
\begin{small}
\begin{equation}\label{eq5}
\begin{gathered}
  {{{V_j}} \mathord{\left/
 {\vphantom {{{V_j}} {{k_{{{jV} \mathord{\left/
 {\vphantom {{jV} I}} \right.
 \kern-\nulldelimiterspace} I}}}{K_{j1}}}}} \right.
 \kern-\nulldelimiterspace} {{k_{{{jV} \mathord{\left/
 {\vphantom {{jV} I}} \right.
 \kern-\nulldelimiterspace} I}}}{K_{j1}}}} = {f_{jx}} + {f_{jy}}{\vartheta _{jz}} - {f_{jz}}{\vartheta _{jy}} + {K_{j0}} + {K_{j2}}{f_{jx}}^2 \hfill \\
   + {K_{j5}}{f_{jy}}^2 + {K_{j7}}{f_{jz}}^2 + {K_{j6}}{f_{jx}}{f_{jy}} + {K_{j4}}{f_{jx}}{f_{jz}} + {K_{j8}}{f_{jy}}{f_{jz}} \thickspace. \hfill \\
\end{gathered}
\end{equation}
\end{small}
To simplify  derivation of the RAGG model, we can rewrite  Eq.\eqref{eq5} as:
\begin{equation}\label{eq6}
\begin{gathered}
  {V_j} = {k_{j1}}{f_{jx}} + {\theta _{jz}}{f_{jy}} - {\theta _{jy}}{f_{jz}} + {k_{j0}} + {k_{j2}}f_{jx}^2 \hfill \\
   + {k_{j5}}f_{jy}^2 + {k_{j7}}f_{jz}^2 + {k_{j4}}{f_{jx}}{f_{jz}} + {k_{j6}}{f_{jx}}{f_{jy}} + {k_{j8}}{f_{jy}}{f_{jz}} \thickspace, \hfill \\
\end{gathered}
\end{equation}
${{k}_{j1}}$, ${{k}_{j0}}$, ${{k}_{j2}}$, ${{k}_{j4}}$, ${{k}_{j5}}$, ${{k}_{j6}}$, ${{k}_{j7}}$, ${{k}_{j8}}$, ${{\theta }_{jy}}$, ${{\theta }_{jz}}$ are given:
\begin{equation}\label{eq6a}
\begin{gathered}
  {k_{j1}} = {k_{jV/I\:}}{K_{j1}},{k_{j0}} = {k_{j1}}{K_{j0}},{k_{j2}} = {k_{j1}}{K_{j2}} \hfill \\
  {k_{j4}} = {k_{j1}}{K_{j4}},{k_{j5}} = {k_{j1}}{K_{j5}},{k_{j6}} = {k_{j1}}{K_{j6}} \hfill \\
  {k_{j7}} = {k_{j1}}{K_{j7}},{k_{j8}} = {k_{j1}}{K_{j8}},{\theta _{jy}} = {k_{j1}}{\vartheta _{jy}} \hfill \\
  {\theta _{jz}} = {k_{j1}}{\vartheta _{jz}} \thickspace.\hfill \\
\end{gathered}
\end{equation}
Where $V_j$ is the voltage output of  accelerometer $A_j$; ${{f}_{jx}},{{f}_{jy}},{{f}_{jz}}$ are the applied specific forces in the directions of the x-, y-, and z-axes, respectively, in the accelerometer nominal frame of the actual mounting position; ${{k}_{j1}}$ is the linear scale factor (in units of V/g); ${{k}_{j0}}$ is the null bias (in units of V); and ${{k}_{j2}}$, ${{k}_{j4}}$, ${{k}_{j5}}$, ${{k}_{j6}}$, ${{k}_{j7}}$, ${{k}_{j8}}$ are the second-order error coefficients (in units of V/g$^2$). From Eq.\eqref{eq6a}, the current to voltage gain mismatch will directly result in the scale factor imbalance.

\subsubsection{ Specific force in the accelerometer nominal frame of the actual mounting position }
\begin{figure}[!t]\centering
	\includegraphics[width=0.4\textwidth]{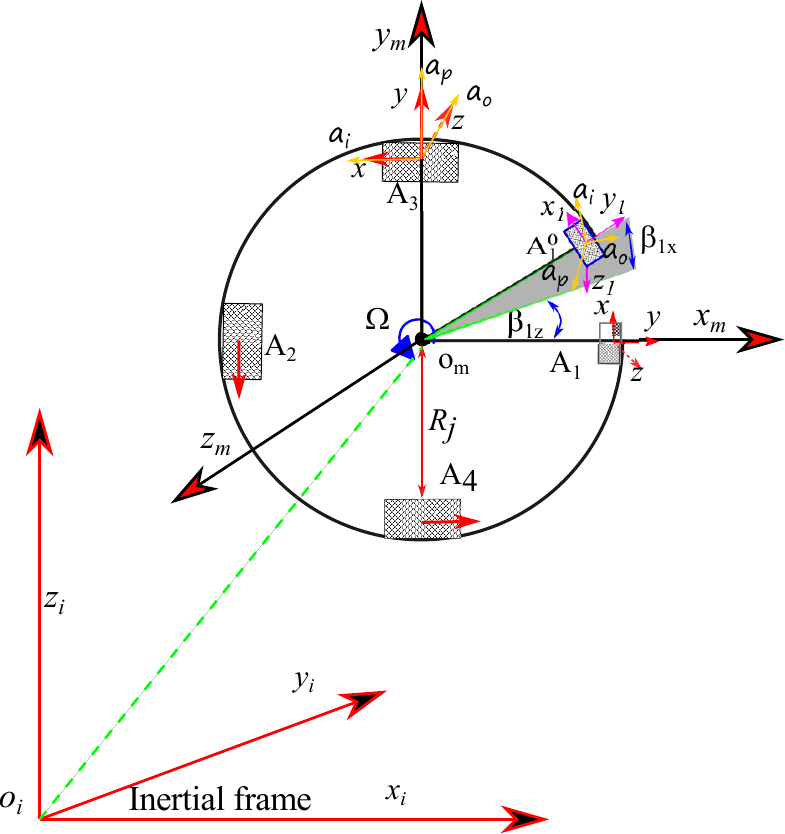}
	\caption{The position vector of the RAGG in the process of moving base gravity gradiometry.}\label{fig2}
\end{figure}

RAGG accelerometers are of the force re-balance type. The measurement is specific force, in other words, the difference between gravitational acceleration and inertial acceleration. Fig.\ref{fig2} illustrates the position vector of the RAGG in the process of moving base gravity gradiometry.
We choose the geocentric inertial coordinate system as the inertial frame, and denote the specific force measured by accelerometer $A_j$ as ${{\bm{f}}_{j}}$:
\begin{equation}\label{eq7}
{{\bm{f}}_j} = {{\bm{a}}_{ji}} - {{\bm{a}}_{gj}}\thickspace,
\end{equation}
where ${{\bm{a}}_{ji}}$ and ${{\bm{a}}_{gj}}$ represent the inertial acceleration and gravitational acceleration of  accelerometer ${{A}_{j}}$, respectively.
The inertial acceleration is the second derivative of the position vector of accelerometer $A_j$ with respect to the inertial frame:
\begin{equation}\label{eq8}
{{\bm{a}}_{ji}} = {\left. {\frac{{d{\bm{r}}_{{o_i}{A_j}}^2}}{{dt}}} \right|_i}{\rm{ = }}{\left. {\frac{{d{\bm{r}}_{{o_i}{o_m}}^2}}{{d{t^2}}}} \right|_i} + {\left. {\frac{{d{\bm{r}}_{{o_m}{A_j}}^2}}{{d{t^2}}}} \right|_i} \thickspace.
\end{equation}
Where ${{\bm{r}}_{{{o}_{i}}{{A}_{j}}}}$ denotes the position vector from the origin of the inertial frame to  accelerometer ${{A}_{j}}$; ${{\bm{r}}_{{{o}_{i}}{{o}_{m}}}}$ denotes the position vector from the origin of the inertial frame to the center of the disc; and ${{\bm{r}}_{{{o}_{m}}{{A}_{j}}}}$ denotes the position from the origin of the RAGG measurement frame to accelerometer ${{A}_{j}}$. The second derivative of ${{\bm{r}}_{{{o}_{m}}{{A}_{j}}}}$ with respect to the inertial frame is given by:
\begin{equation}\label{eq9}
{\left. {\frac{{d{\bm{r}}_{{o_m}{A_j}}^2}}{{d{t^2}}}} \right|_i} = {{\bm{\dot \omega }}_{im}}                   \times                   {{\bm{r}}_{{o_m}{A_j}}} + {{\bm{\omega }}_{im}}                   \times                   \left( {{{\bm{\omega }}_{im}}                   \times                   {{\bm{r}}_{{o_m}{A_j}}}} \right)\thickspace,
\end{equation}
where ${\bm{\omega }_{im}}$ represents the angular velocity of the RAGG with respect to the inertial frame; ${{\dot{\bm{\omega }}}_{im}}$ represents the angular acceleration of the RAGG with respect to the inertial frame. When mass is far enough away from the RAGG, the gravitational acceleration of  accelerometer ${{A}_{j}}$ is a linear approximation of the gravitational acceleration and gravitational gradient tensor at the center of the disc:
\begin{equation}\label{eq10}
{{{\bm{a}}_{gj}} = {{\bm{a}}_{gm}} + {\bm{\Gamma }}     \cdot {{\bm{r}}_{{o_m}{A_j}}}} \thickspace,
\end{equation}
where $\bm{\Gamma }$ denotes the gravitational gradient tensor at the center of the rotating disc.
 Substituting Eqs.\eqref{eq10}, \eqref{eq9}, and \eqref{eq8} into Eq.\eqref{eq7}, we get
\begin{equation}\label{eq11}
\begin{array}{*{20}{l}}
{{{\bm{f}}_j} = {{\bm{f}}_{cmm}} + {{{\bm{\dot \omega }}}_{im}}                   \times                   {{\bm{r}}_{{o_m}{A_j}}} + {{\bm{\omega }}_{im}}                   \times                   \left( {{{\bm{\omega }}_{im}}                   \times                   {{\bm{r}}_{{o_m}{A_j}}}} \right) - {\bm{\Gamma }}                   \cdot {{\bm{r}}_{{o_m}{A_j}}}}\thickspace,\\
{{{\bm{f}}_{cmm}} = {{\left. {\frac{{d{\bm{r}}_{{o_i}{o_m}}^2}}{{d{t^2}}}} \right|}_i} - {{\bm{a}}_{gm}}} \thickspace,
\end{array}
\end{equation}
where ${{\bm{f}}_{cmm}}$ is the specific force at the center of the disc, which is a common mode acceleration component for RAGG accelerometers. We can calculate the specific forces of  accelerometer ${{A}_{j}}$ in the directions of the x-, y-, and z-axes in the accelerometer nominal frame of the actual mounting position, ${{f}_{jx}},{{f}_{jy}},{{f}_{jz}}$, respectively, by:
\begin{equation}\label{eq12}
{\begin{array}{*{20}{l}}
  {{f_{jx}} = {{\bm{f}}_j}                   \cdot {{\bm{\tau }}_{jx}}} \thickspace,\\
  {{f_{jy}} = {{\bm{f}}_j}                   \cdot {{\bm{\tau }}_{jy}}} \thickspace,\\
  {{f_{jz}} = {{\bm{f}}_j}                   \cdot {{\bm{\tau }}_{jz}}}\thickspace,
\end{array}}
\end{equation}
where $\bm{\tau }_{jx}$, $\bm{\tau }_{jy}$, and $\bm{\tau }_{jz}$ are unit vectors of the accelerometer nominal frame of the actual mounting position in the directions of the x-, y-, and z-axes.
Writing the specific force, the angular velocity, and the angular acceleration of the RAGG with respect to the inertial frame in coordinate form gives:
\begin{equation}\label{eq13}
\begin{array}{*{20}{l}}
{{{\bm{f}}_{cmm}} = {{\left[ {{a_x},{a_y},{a_z}} \right]}^T}}\thickspace,\\
{{{\bm{\omega }}_{im}} = {{\left[ {{\omega _{imx}},{\omega _{imy}},{\omega _{imz}}} \right]}^T}}\thickspace,\\
{{{{\bm{\dot \omega }}}_{im}} = {{\left[ {{{ \omega }_{imax}},{{\omega }_{imay}},{{ \omega }_{imaz}}} \right]}^T}}\thickspace.
\end{array}
\end{equation}
Writing the gravitational gradient tensor at the center of the disc in coordinate form gives:
\begin{equation}\label{eq14}
\bm{\Gamma }  = \left[ {\begin{array}{*{20}{c}}
{{\Gamma _{xx}}}&{{\Gamma _{xy}}}&{{\Gamma _{xz}}}\\
{{\Gamma _{xy}}}&{{\Gamma _{yy}}}&{{\Gamma _{yz}}}\\
{{\Gamma _{xz}}}&{{\Gamma _{yz}}}&{{\Gamma _{zz}}}
\end{array}} \right]\thickspace.
\end{equation}
Based on the configuration of the RAGG mentioned in section \ref{sec2.1},
 we can easily get the coordinates of the vectors ${\bm{r}}_{{o_m}{A_j}}$, $\bm{\tau }_{jx}$, $\bm{\tau }_{jy}$, and $\bm{\tau }_{jz}$ .
 Substituting Eqs.\eqref{eq13} and \eqref{eq14} into Eq.\eqref{eq12}, we can calculate the specific force of accelerometers
$A_1$ $\sim$ $A_4$: ${f_{1x}}$, ${f_{1y}}$, ${f_{1z}}$, ${f_{2x}}$, ${f_{2y}}$, ${f_{2z}}$, ${f_{3x}}$, ${f_{3y}}$, ${f_{3z}}$, ${f_{4x}}$, ${f_{4y}}$, ${f_{4z}}$.

\subsubsection{The RAGG analytical model}
We have calculated the specific forces of the four accelerometers: ${f_{1x}}$, ${f_{1y}}$, ${f_{1z}}$, ${f_{2x}}$, ${f_{2y}}$, ${f_{2z}}$, ${f_{3x}}$, ${f_{3y}}$, ${f_{3z}}$, ${f_{4x}}$, ${f_{4y}}$, and ${f_{4z}}$.
Substituting these specific forces into Eq.(\ref{eq6}), we can calculate the output of the four accelerometers. Let $V_1$, $V_2$, $V_3$, and $V_4$ respectively represent the output voltages of the four accelerometers; the output of the RAGG before demodulation is then given by:
\begin{equation}\label{eq15}
G_{out} = {V_1} + {V_2} - {V_3} - {V_4}\thickspace.
\end{equation}
To simplify the description, the notations ${{T}_{1}}                  \sim {{T}_{6}}$ and ${{S}_{1}}                  \sim {{S}_{5}}$ are adopted in RAGG analytic model:
\begin{equation}\label{eq16}
\begin{array}{l}
\left\{ \begin{array}{l}
{T_1} = 0.5({T_{xx}} - {T_{yy}} + \omega _{imy}^2 - \omega _{imx}^2)\\
{T_2} =  - {T_{xy}} + {\omega _{imx}}{\omega _{imy}}\\
{T_3} = ({T_{yz}} - {\omega _{imy}}{\omega _{imz}} - {\omega _{i\max }})\\
{T_4} = ({T_{xz}} - {\omega _{imx}}{\omega _{imz}} + {\omega _{imay}})\\
{T_5} = 0.5({T_{xx}} + {T_{yy}} + \omega _{im}^2 + \omega _{imz}^2)\\
{T_6} = {\omega _{imaz}}
\end{array} \right.
\\
\left\{ \begin{array}{l}
{S_1} = {T_{xz}} - {\omega _{imx}}{\omega _{imz}} - {\omega _{imay}}\\
{S_2} = {\rm{ }}{T_{yz}} - {\omega _{imy}}{\omega _{imz}} + {\omega _{i\max }}\\
{S_3} = \omega _{imx}^2 + \omega _{imz}^2 + {T_{yy}}\\
{S_4} = \omega _{imx}^2 + \omega _{imy}^2 + {T_{zz}}\\
{S_5} =  - {T_{xy}} + {\omega _{imx}}{\omega _{imy}} + {\omega _{imaz}}
\end{array} \right.
\end{array}
\end{equation}
It's worth noting that  ${{T}_{1}}                  \sim {{T}_{6}}$ and ${{S}_{1}}                  \sim {{S}_{6}}$ are consists of gravitational gradients ($\Gamma_{xx}$, $\Gamma_{xy}$, etc.), centrifugal gradients ($\omega _{imx}^2 - \omega _{imy}^2$, ${\omega _{imx}}{\omega _{imy}}$, etc.) and angular accelerations (${{ \omega }_{imax}}$, ${{ \omega }_{imay}}$, ${{\omega }_{imaz}}$).
As the magnitude of centrifugal gradients and angular accelerations are much larger than gravitational gradients, in the error analysis, we can treat ${{T}_{1}}  \sim {{T}_{6}}$ and ${{S}_{1}}   \sim {{S}_{6}}$ as angular motion of a RAGG.
Expanding the Eq.\eqref{eq15} and collecting like terms, yields:
\begin{equation}\label{eq17}
\begin{array}{l}
{G_{out}} = A_{4\Omega }^s\sin 4\Omega t + A_{4\Omega }^c\cos 4\Omega t + A_{3\Omega }^s\sin 3\Omega t
\\ [2pt]
 + A_{3\Omega }^c\cos 3\Omega t + A_{2\Omega }^s\sin 2\Omega t + A_{2\Omega }^c\cos 2\Omega t\\[2pt]
 + A_\Omega ^s\sin \Omega t + A_\Omega ^c\cos \Omega t + {A_0} \thickspace.
\end{array}
\end{equation}
The Eq.\eqref{eq17} is RAGG analytical model; where $A_{4\Omega }^s$, $A_{4\Omega }^c$, $A_{3\Omega }^s$, $A_{3\Omega }^c$, $A_{2\Omega }^s$, $A_{2\Omega }^c$, $A_{\Omega }^s$, and $A_{\Omega }^c$ are the amplitudes of sin$4\Omega t$, cos$4\Omega t$, sin$3\Omega t$, cos$3\Omega t$, sin$2\Omega t$, cos$2\Omega t$, sin$\Omega t$, and cos$\Omega t$.
$A_{4\Omega }^s$, $A_{4\Omega }^c$, $A_{3\Omega }^s$, $A_{3\Omega }^c$, $A_{2\Omega }^s$, and $A_{2\Omega }^c$ are given:
\begin{align}\label{eq18}
A_{4\Omega }^c & = 0.5D_{({k_5} - {k_2}){R^2}}^{1234}({T_1}^2 - {T_2}^2) - D_{{k_6}{R^2}}^{1234}{T_1}{T_2} \notag\\[3.5pt]
A_{4\Omega }^s & =  - D_{({k_5} - {k_2}){R^2}}^{1234}{T_1}{T_2} - 0.5D_{{k_6}{R^2}}^{1234}({T_1}^2 - {T_2}^2)\notag\\[3.5pt]
A_{3\Omega }^s & = 0.5({T_1}{T_4} + {T_2}{T_3})D_{{k_4}{R^2},{k_8}{R^2}}^{12,34}\notag \\[3.5pt] &
 + 0.5({T_1}{T_3}{\rm{ }} - {\rm{ }}{T_2}{T_4}){\rm{ }}D_{ - {k_8}{R^2},{k_4}{R^2}}^{12,34}\notag \\[3.5pt] &
 - {\rm{ }}({T_2}{a_x} - {T_1}{a_y})D_{({k_2} - {k_5})R,{k_6}R}^{12,34}\notag \\[3.5pt] &
 + {\rm{ }}({T_1}{a_x} + {\rm{ }}{T_2}{a_y}){\rm{ }}D_{{k_6}R,({k_5} - {k_2})R}^{12,34}\notag \\[3.5pt]
A_{3\Omega }^c& = 0.5({T_1}{T_4} + {T_2}{T_3}){\rm{ }}D_{ - {k_8}{R^2},{k_4}{R^2}}^{12,34}\notag \\[3.5pt] &
 - 0.5({T_1}{T_3}{\rm{ }} - {\rm{ }}{T_2}{T_4})D_{{k_4}{R^2},{k_8}{R^2}}^{12,34}\notag \\[3.5pt] &
{\rm{ +  }}({T_2}{a_x} - {T_1}{a_y}){\rm{ }}D_{{k_6}R,({k_5} - {k_2})R}^{12,34}\notag \\[3.5pt] &
 + {\rm{ }}({T_1}{a_x} + {\rm{ }}{T_2}{a_y})D_{({k_2} - {k_5})R,{k_6}R}^{12,34}\notag \\[3.5pt]
A_{2\Omega }^s &= {T_1}(\sum {_{{k_1}R}} {\rm{ }} + 2\sum {_{({\beta _z}{\theta _z} + {\beta _x}{\beta _z}{\theta _y})R}} ) + 2{T_1}{T_6}\sum {_{{k_2}{R^2}}} \notag \\[3.5pt] &
 - {T_1}{T_5}\sum {_{{k_6}{R^2}}}  - {T_1}{a_z}\sum {_{{k_4}R}}  + {T_2}\sum {_{({\vartheta _Z} - 2{\beta _Z} + {\beta _x}{\vartheta _y}){k_1}R}}
 \notag \\[3.5pt] &
 - 2{T_2}{T_5}\sum {_{{k_5}{R^2}}}  + {T_2}{T_6}\sum {_{{k_6}{R^2}}}  - {a_z}{T_2}\sum {_{{k_8}R}}
 \notag \\[3.5pt] &
 + {T_3}\left( {0.5{a_x}\sum {_{{k_8}R}}  + 0.5{a_y}\sum {_{{k_4}R}} } \right) + \sum {_{{k_7}{R^2}}} {T_3}{T_4}
 \notag \\[3.5pt] &
 - {T_4}\left( {0.5{a_x}\sum {_{{k_4}R}}  - 0.5{a_y}\sum {_{{k_8}R}} } \right) + {a_x}{a_y}\sum {_{{k_5} - {k_2}}}
  \notag \\[3.5pt] &
 - 0.5\sum {_{{k_6}}} \left( {{a_x}^2 - {a_y}^2} \right) \notag \\[3.5pt]
A_{2\Omega }^c &= {T_2}(\sum {_{{k_1}R}} {\rm{ }} + 2\sum {_{({\beta _z}{\theta _z} + {\beta _x}{\beta _z}{\theta _y})R}} ) + 2{T_2}{T_6}\sum {_{{k_2}{R^2}}}  \notag \\[3.5pt] &
 - {T_2}{T_5}\sum {_{{k_6}{R^2}}}  - {T_2}{a_z}\sum {_{{k_4}R}}  - {T_1}\sum {_{({\vartheta _Z} - 2{\beta _Z} + {\beta _x}{\vartheta _y}){k_1}R}}
  \notag \\[3.5pt] &
 + 2{T_1}{T_5}\sum {_{{k_5}{R^2}}}  - {T_1}{T_6}\sum {_{{k_6}{R^2}}}  + {T_1}{a_z}\sum {_{{k_8}R}}
  \notag \\[3.5pt] &
 + {T_3}\left( {0.5{a_x}\sum {_{{k_4}R}}  - 0.5{a_y}\sum {_{{k_8}R}} } \right) + {a_x}{a_y}\sum {_{{k_6}}}
  \notag \\[3.5pt] &
 + {T_4}\left( {0.5{a_x}\sum {_{{k_8}R}}  + 0.5{a_y}\sum {_{{k_4}R}} } \right)
  \notag \\[3.5pt] &
 + 0.5\left( {{a_x}^2 - {a_y}^2} \right)\sum {_{{k_5} - {k_2}}}  + 0.5\left( {{T_4}^2 - {T_3}^2} \right)\sum {_{{k_7}{R^2}}}
 \notag \\
\end{align}
$A_\Omega ^s$, $A_{\Omega }^c$, $A_0$ are given:
\begin{align}\label{eq19}
A_\Omega ^s{\rm{ }} &= {\rm{ }}{S_1}D_{({\beta _X} + \beta _X^2{\beta _Z}{\vartheta _Y}){k_1}R,({\beta _X}{\beta _Z} - \beta _X^2{\vartheta _Y}){k_1}R}^{12,34} \notag \\[3.5pt] &
 + {S_2}D_{({\beta _X}{\beta _Z} - \beta _X^2{\vartheta _Y}){k_1}R, - ({\beta _X} + \beta _X^2{\beta _Z}{\theta _Y}){k_1}R}^{12,34} - D_{{k_8},{k_4}}^{12,34}{a_y}{a_z}{\rm{ }} \notag \\[3.5pt] &
 + {\rm{ }}{a_x}D_{ - (1 + {\beta _z}{\vartheta _z} + {\beta _X}{\beta _Z}{\vartheta _Y}){k_1},({\vartheta _Z} - {\beta _Z} + {\beta _X}{\vartheta _Y}){k_1}}^{12,34} - D_{({k_2} + {k_5})R}^{34}{T_1}{a_x} \notag \\[3.5pt] &
 + {\rm{ }}D_{({k_5} + {k_2})R}^{12}{T_2}{a_x} + {\rm{ }}D_{{k_6}R, - 2{k_5}R}^{12,34}{T_5}{a_x} + D_{ - 2{k_2}R,{k_6}R}^{12,34}{T_6}{a_x} \notag \\[3.5pt] &
 + {\rm{ }}{a_y}D_{({\vartheta _z} - {\beta _z} + {\beta _X}{\vartheta _Y}){k_1},(1 + {\beta _z}{\vartheta _z} + {\beta _X}{\beta _Z}{\vartheta _Y}){k_1}}^{12,34} + {\rm{ }}D_{({k_2} + {k_5})R}^{12}{T_1}{a_y} \notag \\[3.5pt] &
 + {\rm{ }}D_{({k_5} + {k_2})R}^{34}{T_2}{a_y} - D_{2{k_5}R,{k_6}R}^{12,34}{T_5}{a_y} + {\rm{ }}D_{{k_6}R,2{k_2}R}^{12,34}{T_6}{a_y} \notag \\[3.5pt] &
 - \left( {2D_{{k_7}R}^{12}{T_3}{\rm{ }} + 2D_{{k_7}R}^{34}{T_4}} \right){a_z} + {\rm{ }}D_{{k_4}, - {k_8}}^{12,34}{a_x}{a_z} \notag \\[3.5pt] &
 - 2D_{{\beta _X}{\theta _Z}R,{\beta _X}{\beta _Z}{\theta _Z}R}^{12,34}{\omega _{i\max }} + 2D_{ - {\beta _X}{\beta _Z}{\theta _Z}R,{\beta _X}{\theta _Z}R}^{12,34}{\omega _{imay}} \notag \\[3.5pt] &
 + {\rm{ }}D_{{k_4}{R^2}}^{12}\left( {0.5{T_1}{T_4}{\rm{ }} - {\rm{ }}0.5{T_2}{T_3}{\rm{ }} + {\rm{ }}{T_3}{T_6}} \right) \notag \\[3.5pt] &
 + {\rm{ }}D_{{k_4}{R^2}}^{34}\left( {0.5{T_1}{T_3}{\rm{ }} + {\rm{ }}0.5{T_2}{T_4}{\rm{ }} + {\rm{ }}{T_4}{T_6}} \right) \notag \\[3.5pt] &
 - {\rm{ }}D_{{k_8}{R^2}}^{34}\left( {0.5{T_1}{T_4}{\rm{ }} - 0.5{T_2}{T_3}{\rm{ }} + {\rm{ }}{T_4}{T_5}} \right) \notag \\[3.5pt] &
 + {\rm{ }}D_{{k_8}{R^2}}^{12}\left( {0.5{T_1}{T_3}{\rm{ }} + {\rm{ }}0.5{T_2}{T_4}{\rm{ }} - {\rm{ }}{T_3}{T_5}} \right) \notag \\[3.5pt] &
 - D_{{\theta _Y}R,{\beta _Z}{\theta _Y}R}^{12,34}{T_3}{\rm{ }} + D_{{\beta _Z}{\theta _Y}R, - {\theta _Y}R}^{12,34}{T_4} \notag \\[3.5pt]
 A_\Omega ^c &= {S_1}D_{({\beta _X}{\beta _Z} - \beta _X^2{\vartheta _y}){k_1}R, - ({\beta _X} + \beta _X^2{\beta _Z}{\vartheta _Y}){k_1}R}^{12,34}
\notag \\[3.5pt] &
 - {\rm{ }}{S_2}D_{({\beta _X} + \beta _X^2{\beta _Z}{\vartheta _Y}){k_1}R,({\beta _X}{\beta _Z} - \beta _X^2{\vartheta _y}){k_1}R}^{12,34} - D_{{k_8},{k_4}}^{12,34}{a_x}{a_z}{\rm{ }}
\notag \\[3.5pt] &
 + {a_x}D_{({\vartheta _z} - {\beta _z} + {\beta _X}{\vartheta _Y}){k_1},(1 + {\beta _z}{\vartheta _z} + {\beta _X}{\beta _Z}{\vartheta _Y}){k_1}}^{12,34} - {\rm{ }}D_{({k_2} + {k_5})R}^{12}{T_1}{a_x}
\notag \\[3.5pt] &
 - D_{({k_2} + {k_5})R}^{34}{T_2}{a_x} - D_{2{k_5}R,{k_6}R}^{12,34}{T_5}{a_x} + {\rm{ }}D_{{k_6}R,2{k_2}R}^{12,34}{T_6}{a_x}
\notag \\[3.5pt] &
 + {a_y}D_{(1 + {\beta _z}{\vartheta _z} + {\beta _X}{\beta _Z}{\vartheta _Y}){k_1}, - ({\vartheta _z} - {\beta _z} + {\beta _X}{\vartheta _Y}){k_1}}^{12,34} - D_{({k_2} + {k_5})R}^{34}{T_1}{a_y}
\notag \\[3.5pt] &
 + D_{({k_2} + {k_5})R}^{12}{T_2}{a_y} + D_{ - {k_6}R,2{k_5}R}^{12,34}{T_5}{a_y}{\rm{ + }}D_{2{k_2}R, - {k_6}R}^{12,34}{T_6}{a_y}
\notag \\[3.5pt] &
 + {\rm{ }}\left( {2D_{{k_7}R}^{34}{T_3} - {\rm{ }}2D_{{k_7}R}^{12}{T_4}} \right){a_z} + {\rm{ }}D_{ - {k_4},{k_8}}^{12,34}{a_y}{a_z}
 \notag \\[3.5pt] &
 + 2D_{ - {\beta _X}{\beta _Z}{\theta _Z}R,{\beta _X}{\theta _Z}R}^{12,34}{\omega _{i\max }} + 2D_{{\beta _X}{\theta _Z}R,{\beta _X}{\beta _Z}{\theta _Z}R}^{12,34}{\omega _{imay}}
\notag \\[3.5pt] &
{\rm{ }} + D_{{k_4}{R^2}}^{12}{\rm{ }}(0.5{T_1}{T_{3{\rm{ }}}} + {\rm{ }}0.5{T_2}{T_4}{\rm{ }} + {\rm{ }}{T_4}{T_6})
\notag \\[3.5pt] &
 - D_{{k_8}{R^2}}^{12}(0.5{T_1}{T_4}{\rm{ }} - {\rm{ }}0.5{T_2}{T_3}{\rm{ }} + {\rm{ }}{T_4}{T_5}){\rm{  }}\notag \\[3.5pt] &
{\rm{ }} - D_{{k_4}{R^2}}^{34}(0.5{T_1}{T_4}{\rm{ }} - {\rm{ }}0.5{T_2}{T_3}{\rm{ }} + {\rm{ }}{T_3}{T_6}){\rm{ }}\notag \\[3.5pt] &
 - D_{{k_8}{R^2}}^{34}(0.5{T_1}{T_3}{\rm{ }} + {\rm{ }}0.5{T_2}{T_4}{\rm{ }} - {\rm{ }}{T_3}{T_5}){\rm{  }}\notag \\[3.5pt] &
{\rm{ }} - {\rm{ }}D_{{\theta _Y}R,{\beta _Z}{\theta _Y}R}^{12,34}{T_4}{\rm{  }} - D_{{\beta _Z}{\theta _Y}R, - {\theta _Y}R}^{12,34}{T_3} \notag \\[3.5pt]
{A_0} & = 0.5D_{{k_2} + {k_5}}^{1234}({a_x}^2 + {a_y}^2)\notag \\[3.5pt] &
 + 0.5(D_{{k_8}R}^{1234}{T_4} - D_{{k_4}R}^{1234}{T_3}){a_x}\notag \\[3.5pt] &
 + 0.5(D_{{k_4}R}^{1234}{T_4} + D_{{k_8}R}^{1234}{T_3}){a_y}\notag \\[3.5pt] &
 + (D_{{\theta _Y} - {\beta _X}{\theta _Z}}^{1234} - D_{{k_4}R}^{1234}{T_6} + D_{{k_8}R}^{1234}{T_5}){a_z}\notag \\[3.5pt] &
 + D_{{k_7}}^{1234}{a_z}^2 + 0.5D_{2{\beta _X}{\beta _X}{\theta _Z}R + {\theta _y}{\beta _x}R}^{1234}{S_4}\notag \\[3.5pt] &
 - D_{{\beta _X}{\theta _Y}R}^{1234}\omega _{im}^2 + D_{{k_2}{R^2}}^{1234}(0.5{T_1}^2 + 0.5{T_2}^2 + {T_6}^2)\notag \\[3.5pt] &
 + D_{{k_5}{R^2}}^{1234}(0.5{T_1}^2 + 0.5{T_2}^2 + {T_5}^2)\notag \\[3.5pt] &
 + 0.5D_{{k_7}{R^2}}^{1234}({T_3}^2 + {T_4}^2) - D_{{k_6}{R^2}}^{1234}{T_5}{T_6}\notag \\[3.5pt] &
 - D_{{\theta _Z}R}^{1234}{T_5} + D_{{k_1}R}^{1234}{T_6} + D_{{k_0}}^{1234}
\end{align}

In Eqs.\eqref{eq18}, \eqref{eq19}, the notation $D^{12}_{subscript}$ represents the imbalance terms denoted by subscript between accelerometers $A_1$ and $A_2$; the notation $D^{34}_{subscript}$ represents the imbalance terms denoted by subscript between accelerometers $A_3$ and $A_4$; the notation $D^{12,34}_{subscript1,subscript2}$ is the sum of $D^{12}_{subscript1}$ and $D^{34}_{subscript2}$; the notation $D^{1234}_{subscript}$ represents the imbalance terms denoted by subscript between two pairs accelerometers $A_1$, $A_2$ and $A_3$, $A_4$.
For example $D^{12}_{k1R}$ is the imbalance term of $k_{1}R$ between accelerometers $A_1$ and $A_2$, that is $D^{12}_{k1R}=k_{11}R_1-k_{22}R_2$;
$D^{1234}_{k4R}$ is the imbalance term of $k_{4}R$ between two pairs accelerometers, that is $D^{1234}_{k4R}=k_{14}R_1+k_{24}R_2-k_{34}R_{3}-k_{44}R_{4}$. Similarly the notation $\sum{_{subscript}}$ represents the sum of the four accelerometers of the terms denoted by subscript; for example, $\sum{_{k_{1}R}}$ is the sum of four accelerometers of $k_{1}R$, that is, $\sum{_{k_{1}R}}=k_{11}R_1+k_{21}R_2+k_{31}R_3+k_{41}R_4$.
If the accelerometers of the RAGG are perfectly mounted, the accelerometer linear scale factors are balanced, and the accelerometer seconder-order error coefficients are zero, the output of the RAGG is:
\begin{equation}\label{eq30}
G_{out}^* = \sum {_{{k_1}R}} {T_1}\sin 2\Omega t + \sum {_{{k_1}R}} {T_2}\cos 2\Omega t \thickspace,
\end{equation}
where $G_{out}^*$ is the ideal output of the RAGG before demodulation. Let $k_{ggi}$ denote the scale factors of the RAGG,
 thus,  $k_{ggi}=\sum {_{{k_1}R}}$. Since $k_{ggi}$ also is a error propagation coefficient of centrifugal gradient, the parameters of $G_{out}$, $\sum {_*}$, $D^{12}_*$, $D^{34}_*$, etc., are error propagation coefficients.
 Actually, in moving base gravity gradiometry, some error propagation coefficients have little effect on the sensitivity of RAGG and could be neglected.
So, next we will further simplify the RAGG analytical model.

\subsection{Simplified RAGG Analytical Model}
\subsubsection{Simplifying the RAGG analytical model}
In the RAGG analytical model, ${{T}_{1}}$ $\sim$ ${{T}_{6}}$ and ${{S}_{1}}$ $\sim$ ${{S}_{6}}$ refer to the angular motion of a RAGG, while $a_x$, $a_y$, and $a_z$ refer to the linear motion of a RAGG. The error propagation coefficients are permutations of factors such as the misalignment angle, the scale factors, and the second-order error coefficients.
The error propagation coefficients transfer the linear motion and angular motion into the output of the RAGG, causing measurement errors.
Assuming that the RAGG sensitivity is~1 Eo and the nominal distance from the RAGG accelerometer to the center of the disc is 0.1~m, the accelerometer mounting errors ${\beta _{jx}}$, ${\beta _{jz}}$, ${\vartheta _{jy}}$, and ${\vartheta _{jz}}$ are of the order of ${10^{-4}}$~rad, the accelerometer linear-scale-factor imbalance is of the order of ${10^{-4}}$, and the accelerometer second-order error coefficients are of the order of $10^{-6}$~g/g$^2$. Under these conditions, we will calculate the critical conditions for the linear and angular motions such that the error terms can be neglected. By comparing the critical conditions with those of the actual moving-base gravity gradiometry environment, we can determine whether the error terms should be ignored.

In the RAGG analytical model (Eq.\eqref{eq18}\textasciitilde\eqref{eq19}), classified by error sources, all error terms can be divided into six categories:
 coupling error terms concerning second-order error coefficients and angular motion, coupling error terms concerning second-order error coefficients and linear motion, coupling error terms concerning linear scale-factor imbalance and linear motion, coupling error terms concerning second-order error coefficients, linear motion and angular motion, coupling error terms concerning misalignment angle, linear scale factors, and angular motion, coupling error terms concerning misalignment angles, linear scale factors, and linear motion. We simplify the RAGG analytical model by categories.

\begin{compactenum}[a)]
	\item Coupling error terms concerning second-order error coefficients and angular motion.
As mentioned in Section~\ref{sec2.1}, the parameters ${{k}_{jp}}$ ($j=1,2,3,4$; $p=2,4,5,6,7,8$) in the accelerometer model represent the second-order error coefficients ${{k}_{p}}$ of the accelerometer $A_j$. ${{T}_{1}}$ $\sim$ ${{T}_{6}}$ and ${{S}_{1}}$ $\sim$ ${{S}_{6}}$ refer to the angular motion of a RAGG, so the basic coupling error terms concerning the second-order error coefficients and angular motion are of the form ${{k}_{jp}}{{T}_{n_1}}{{T}_{n_2}}{{R_j}^{2}}$ ($n_1, n_2=1,2,3,4,5,6$; $p=2,4,5,6,7,8$). In the analytical model $G_{out}$, the terms $\sum_{k_jR^2}T_{n_1}T_{n_2}$ and $D^{*}_{k_jR^2}T_{n_1}T_{n_2}$ ($n_1, n_2=1,2,3,4,5,6$) consist of ${{k}_{jp}}{{T}_{n_1}}{{T}_{n_2}}{{R_j}^{2}}$ and belong to the coupling error terms concerning the second-order error coefficients and angular motion. The RAGG measurement error $M_{e1}$ due to ${{k}_{jp}}{{T}_{n_1}}{{T}_{n_2}}{{R_j}^{2}}$ can be expressed as
\begin{equation}\label{eq31}
M_{e1} = {{{k_{jp}}{T_{n1}}{T_{n_2}}{R_j^2}} \mathord{\left/
 {\vphantom {{{k_{jp}}{T_{n_1}}{T_{n_2}}{R_j^2}} {{k_{ggi}}}}} \right.
 \kern-\nulldelimiterspace} {{k_{ggi}}}} \thickspace ,
\end{equation}
where ${{k}_{ggi}}$ is the RAGG scale factor. In the accelerometer model mentioned in Section~\ref{sec2.1}, we have  ${{k}_{jp}}={{k}_{j1}}{{K}_{jp}}$, so we obtain
\begin{equation}\label{eq32}
M_{e1} \approx 0.25{K_{jp}}{T_{n_1}}{T_{n_2}}R_j  \thickspace .
\end{equation}
${{T}_{{n_1}}}$ and ${{T}_{{n_2}}}$ are of the same order, and we use $(T_n)_{critical}$ to denote the critical value of $T_{n_1}$ and $T_{n_2}$. To ensure an RAGG sensitivity of 1~Eo, it is reasonable to assume that the error contributed by ${{k}_{jp}}{{T}_{n_1}}{{T}_{n_2}}{{R_j}^{2}}$ is 0.1~Eo. Substituting ${{K}_{jp}} = 10^{-6}$~g/g$^2$, $R_j =0.1$~m, and $M_{e1}=0.1$~Eo into Eq.~\eqref{eq32}, we calculate the critical value $(T_n)_{critical}$  as $2                  \times                  10^8$~Eo. Based on Eq.~\eqref{eq16}, $T_1 \sim T_6$ are the sums of the gravitational gradients, centrifugal gradients, and angular accelerations. The maximum gravitational gradient is of the order of $10^3$~Eo. Actually, in moving-base gravity gradiometry, $T_1\sim T_6$ are principally the centrifugal gradients and angular accelerations caused by the angular motion of the RAGG. Here, we will calculate the critical angular motion of the RAGG according to $(T_n)_{critical}$.

The unit transformations from 1~Eo to angular velocity squared (rad$^2$/s$^2$) and angular acceleration (rad/s$^2$) are given:
\begin{equation*}
\left\{ \begin{array}{l}
1~\rm{Eo} = {10^{-9}}~{{ra{d^2}} \mathord{\left/
 {\vphantom {{ra{d^2}} {{s^2}}}} \right.
 \kern-\nulldelimiterspace} {{s^2}}}\\
1~\rm{Eo} = {10^{-9}}~{{rad} \mathord{\left/
 {\vphantom {{rad} {{s^2}}}} \right.
 \kern-\nulldelimiterspace} {{s^2}}}  \thickspace .
\end{array} \right.
\end{equation*}
Clearly, $T_{n_1}  \le (T_n)_{critical}$ is the condition for neglecting the error terms ${{k}_{jp}}{{T}_{n_1}}{{T}_{n_2}}{{R_j}^{2}}$. That is,
\begin{equation}\label{eq33}
\begin{array}{l}
{T_1} \approx 0.5(\omega _{imy}^2 - \omega _{imx}^2)                   \le {({T_n})_{critical}},\\
{T_2} \approx ({\omega _{imx}}{\omega _{imy}})                   \le {({T_n})_{critical}},\\
{T_3} \approx ({{ \omega }_{imax}} + {\omega _{imy}}{\omega _{imz}})                   \le {({T_n})_{critical}},\\
{T_4} \approx ({\omega _{imx}}{\omega _{imz}} - {{ \omega }_{imay}})                   \le {({T_n})_{critical}},\\
{T_5} \approx 0.5(\omega _{imx}^2 + \omega _{imy}^2 + 2\omega _{imz}^2)                   \le {({T_n})_{critical}},\\
{T_6} \approx {{\omega }_{imaz}}                   \le {({T_n})_{critical}}   \thickspace .
\end{array}
\end{equation}
Assuming that the angular velocity  and  acceleration components  are of the same order, we have
${\omega _{imx}} = {\omega _{imy}} = {\omega _{imz}}$ and ${{ \omega }_{imax}} = {{\omega }_{imay}} = {{\omega }_{imaz}}$. Let ${{\omega }_{critical}}$ and ${{\dot{\omega }}_{critical}}$  represent the critical angular velocity component and the critical angular acceleration component, respectively. Then by Eq.~(\ref{eq33}), we can  calculate the critical value of the angular motion roughly as
\begin{equation}\label{eq34}
 \begin{array}{l}
\omega _{critical} = 0.22~\rm{rad/s} = 12.74^\circ/\rm{s},\\
\dot\omega_{critical} = 0.148~\rm{rad/s}^2 = 8.51^\circ/\rm{s}^2.
\end{array}
\end{equation}
Those harmonic components of the angular motion whose fundamental frequency equal the rotation rate of the rotating disc have considerable impact on the RAGG sensitivity. The angular acceleration is the derivative of the angular velocity. Therefore, we can denote the harmonic components of the angular motion as
\begin{equation}\label{eq35}
\begin{array}{l}
\omega_{k} \left( t \right) = A_{k\Omega }\sin \left( {k\Omega t} \right),\\
\dot \omega_{k} \left( t \right) = k\Omega A_{k\Omega }\cos \left( {k\Omega t} \right) \thickspace ,
\end{array}
\end{equation}
where $\omega_{k} \left( t \right)$ and $\dot \omega_{k} \left( t \right)$ are the $k$th-order harmonic components of angular velocity and angular acceleration, respectively, $\Omega$ is the angular frequency of the rotating disk, and $A_{k\Omega }$ is the magnitude of the $k$th-order harmonic component of angular velocity. Let ${{A}_{k\Omega critical}}$ represent the critical magnitude of the $k$th-order harmonic component of angular velocity. Obviously,
\begin{equation}\label{eq36}
{A_{k\Omega critical}}                   \le \min \{ {\omega _{critical}},{{{{\dot \omega }_{critical}}} \mathord{\left/
 {\vphantom {{{{\dot \omega }_{critical}}} {k\Omega }}} \right.
 \kern-\nulldelimiterspace} {k\Omega }}\}   \thickspace .
 \end{equation}
Substituting $k=1$ and $\Omega=1.57$~rad/s (the frequency of the rotating disc is 0.25~Hz) into Eq.~(\ref{eq36}), we calculate the critical magnitude of the fundamental frequency component as $A_{\Omega critical} = 0.0945$~rad/s = 5.42$^\circ$/h. Based on the above analysis, we list in Table~\ref{tab1} the conditions for neglecting ${k_{jp}}{T_{n_1}}{T_{n_2}}{R_j^2}$ .
\renewcommand{\multirowsetup}{\centering}
\begin{table}[htbp]
\centering
\caption{Conditions for Neglecting Coupling Error Terms Concerning Second-order Error Coefficients and Angular Motion}
\label{tab1}
\begin{tabular}{ll}
\hline \hline
Error terms & Conditions\\
\hline
${k_{jp}}{T_{n_1}}{T_{n_2}}{R_j^2}$ &
$\begin{array}{l}
\omega_{im}                    \le {\omega _{critical}} = {\rm{12}}{\rm{.74}}{{^                   \circ } \mathord{\left/
 {\vphantom {{^                   \circ } s}} \right.
 \kern-\nulldelimiterspace} s}\\
{{\dot \omega_{im} }}                   \le {{\dot \omega }_{critical}}= {\rm{8}}{\rm{.51}}{{^                   \circ } \mathord{\left/
 {\vphantom {{^                   \circ } {{s^2}}}} \right.
 \kern-\nulldelimiterspace} {{s^2}}}\\
{A_\Omega }                   \le {{A_{\Omega critical} }} = {\rm{ }}{{{\rm{5}}{\rm{.42}}{{\rm{4}}^                   \circ }} \mathord{\left/
 {\vphantom {{{\rm{5}}{\rm{.42}}{{\rm{4}}^                   \circ }} s}} \right.
 \kern-\nulldelimiterspace} s}
\end{array}$ \\
\hline \hline
\end{tabular}
\end{table}
In gravity gradiometry, the RAGG is mounted on a stabilized platform that is isolated from high-frequency vibrations by pneumatic mounting pads. It is relatively easy to meet the conditions listed in Table~\ref{tab1}, meaning that the error terms concerning the second-order error coefficients and angular motion ($\sum_{k_jR^2}T_{n_1}T_{n_2}$, $D^{*}_{k_jR^2}T_{n_1}T_{n_2}$) can be neglected.
From the perspective of unit operation of physical quantity, the error terms concerning the second-order error coefficients and angular motion include $T_{n1}T_{n2}$, and $T_{n1}T_{n2}$ will only be coupled to the accelerometer second-order error coefficients, so in Eqs.\eqref{eq18} and \eqref{eq19},
any item containing $T_{n1}T_{n2}$ can be neglected.

 \item Coupling error terms concerning second-order error coefficients and linear motion. We apply $k_{jp}{a}_{n_1}{a}_{n_2}$ ($n_1, n_2 =x,y,z$; $p=2,4,5,6,7,8$) representing the basic coupling error terms concerning the second-order error coefficients and linear motion. In the analytical model $G_{out}$, the terms $D^{*}_{k_{jp}}a_{n_1}a_{n_2}$ and $\sum_{k_{jp}}a_{n_1}a_{n_2}$ consist of $k_{jp}{a}_{n_1}{a}_{n_2}$ and belong to the coupling error terms concerning the second-order error coefficients and linear motion; correspondingly, the measurement errors contributed by $k_{jp}{a}_{n_1}{a}_{n_2}$ can be expressed as
\begin {equation}\label{eq36a}
\begin{array}{l}
{M_{e2}} = {{{k_{jp}}{a_{{n_1}}}{a_{{n_2}}}} \mathord{\left/
 {\vphantom {{{k_{jp}}{a_{{n_1}}}{a_{{n_2}}}} {{k_{ggi}}}}} \right.
 \kern-\nulldelimiterspace} {{k_{ggi}}}}   \thickspace .
\end{array}
\end{equation}
 In the accelerometer, ${{k}_{jp}}={{k}_{j1}}{{K}_{jp}}$, so Eq.~(\ref{eq36a}) becomes
\begin{equation}\label{eq37}
\begin{array}{l}
{M_{e2}} \approx {{0.25{K_p}{a_{{n_1}}}{a_{{n_2}}}} \mathord{\left/
 {\vphantom {{0.25{K_p}{a_{{n_1}}}{a_{{n_2}}}} {{R_j}}}} \right.
 \kern-\nulldelimiterspace} {{R_j}}} \thickspace .
\end{array}
\end{equation}
Let $a_{1critical}$ denotes the critical accelerations of $k_{jp}{a}_{n_1}{a}_{n_2}$. Substituting $M_{e2}=0.1$~Eo and $K_{jp}=10^{-6}$~g into Eq.~(\ref{eq37}), we get $a_{1critical}=2\times 10^{-3}$~g. In moving-base gradiometry, the linear acceleration $a_n$ is of the order of 0.1~g. Therefore, the coupling error terms concerning second-order error coefficients and linear motion cannot be neglected.

 \item Coupling error terms concerning linear scale-factor imbalance and linear motion. We apply $dk_{j1}a_{n}$ ($n=x,y,z$) representing the basic coupling error terms concerning the linear-scale-factor imbalance and linear motion. In the analytical model $G_{out}$, the terms $\sum_{k_{j1}}a_n$ and $D^{*}_{k_{j1}}*a_n$ consist of $dk_{j1}a_{n}$ and belong to the coupling error terms concerning the linear-scale-factor imbalance and linear motion. Correspondingly, the measurement errors contributed by $dk_{j1}a_{n}$ can be expressed as
\begin {equation}\label{eq36aa}
\begin{array}{l}
{M_{e3}} = {{d{k_{j1}}{a_n}} \mathord{\left/
 {\vphantom {{d{k_{j1}}{a_n}} {{k_{ggi}}}}} \right.
 \kern-\nulldelimiterspace} {{k_{ggi}}}}   \thickspace .
\end{array}
\end{equation}
The degree of linear-scale-factor imbalance before online adjustment is of the order of $10^{-4}$, i.e.,  ${d{{k}_{j1}}}/{{{k}_{j1}}}\;={{10}^{-4}}$. $k_{ggi}$ is the RAGG scale factor. so Eq.~\eqref{eq36aa} becomes
\begin{equation}\label{eq37a}
\begin{array}{l}
{M_{e3}} \approx {{{2.5                   \times                   {{10}^{ - 5}}{a_n}} \mathord{\left/
 {\vphantom {{2.5                   \times                   {{10}^{ - 5}}{a_i}} R}} \right.
 \kern-\nulldelimiterspace} R}_j}   \thickspace .
\end{array}
\end{equation}
Let $a_{2critical}$ denotes the critical accelerations of $dk_{j1}a_{n}$, respectively. Substituting $M_{e3}=0.1$~Eo into Eq.~\eqref{eq37a}, we get $a_{2critical}=4\times10^{-8}$~g. In moving-base gradiometry, the linear acceleration $a_n$ is of the order of 0.1~g. Therefore, the coupling error terms concerning the linear-scale-factor imbalance and linear motion cannot be neglected.

\item Coupling error terms concerning second-order error coefficients, linear motion and angular motion.
We denote the basic coupling error terms concerning second-order error coefficients, linear motion, and angular motion as ${{k}_{jp}}{{T}_{n_1}}R_j{{a}_{n_2}}$. In the RAGG analytic model, $D^{*}_{k_{jp}}T_{n_1}R_ja_{n_2}$ and $\sum_{k_{jp}}T_{n_1}R_ja_{n_2}$ are the coupling error terms concerning the second-order error coefficients, linear motion, and angular motion. The measurement error contributed by  ${{k}_{jp}}{{T}_{n_1}}R_j{{a}_{n_2}}$ is
\begin{equation}\label{eq38}
{M_{e4}} = {{{k_{jp}}{T_{{n_1}}}{R_j}{a_{{n_2}}}} \mathord{\left/
 {\vphantom {{{k_{jp}}{T_{{n_1}}}{R_j}{a_{{n_2}}}} {{k_{ggi}}}}} \right.
 \kern-\nulldelimiterspace} {{k_{ggi}}}} \approx 0.25{K_{jp}}{T_{{n_1}}}{a_{{n_2}}}   \thickspace .
 \end{equation}
In gravity gradiometry, the acceleration $a_{n_2}$ is of the order of 0.1~g. To ensure an RAGG sensitivity of 1~Eo and assuming $M_{e4} = 0.1$~Eo, $K_{jp}$=10$^{-6}$~g/g$^2$, and $a_{n_2} = 0.1$~g, based on Eq.~\eqref{eq38}, we obtain the critical value of $T_{n_1}$ for neglecting ${{k}_{jp}}{{T}_{n_1}}R_j{{a}_{n_2}}$, namely $(T_n)_{critical}=4 \times 10^{7}$~Eo. Similar to the previous analysis, the critical angular velocity and angular accelerations are calculated and are listed in Table~\ref{tab2}.
\begin{table}[htbp]
\caption{Conditions for Neglecting Coulping Error Terms Concerning Linear Motion, Angular Motion, and Second-order Error Coefficients}
\label{tab2}       
\centering
\begin{tabular}{ll}
\hline \hline
Error terms & Conditions\\
\hline
${k_{jp}}{T_{n_1}}R_j{a_{n_2}}$&
$\begin{array}{l}
\omega_{im}                   \le {\omega _{critical}} = {{{\rm{5}}{\rm{.73}}{{\rm{4}}^                   \circ }} \mathord{\left/
 {\vphantom {{{\rm{5}}{\rm{.73}}{{\rm{4}}^                   \circ }} s}} \right.
 \kern-\nulldelimiterspace} s}\\
{{\dot \omega }_{im}}                   \le {{{\dot \omega }}_{critial}} = {\rm{1}}{\rm{.72}}{{^                   \circ } \mathord{\left/
 {\vphantom {{^                   \circ } {{s^2}}}} \right.
 \kern-\nulldelimiterspace} {{s^2}}}\\
{A_\Omega }                   \le {{A_{\Omega critical}}} = {\rm{1}}{\rm{.09}}{{^                   \circ } \mathord{\left/
 {\vphantom {{^                   \circ } s}} \right.
 \kern-\nulldelimiterspace} s}
\end{array}$\\
\hline \hline
\end{tabular}
\end{table}
In moving-base gradiometry, it is relatively easy to satisfy the conditions listed in Table~\ref{tab2}, so the coupling error terms concerning second-order error coefficients, linear motion and angular motion can be neglected.
From the perspective of unit operation of physical quantity, the error terms concerning the second-order error coefficients, linear motion and angular motion include $T_{ni}a_{nj}$, and $T_{ni}a_{nj}$ will only be coupled to the accelerometer second-order error coefficients, so in Eqs.\eqref{eq18} and \eqref{eq19}, any item containing $T_{ni}a_{nj}$ can be neglected.

\item Coupling error terms concerning mounting misalignment angle, linear scale factors, and angular motion.
The mounting misalignment angles are $\beta_{jx}$, $\beta_{jz}$, $\vartheta_{jy }$, and $\vartheta_{jz}$. The basic coupling error terms concerning
\lineskiplimit=1pt \lineskip=3.5pt
the misalignment angle, linear scale factors, and angular motion are permutations of the misalignment angles ($\beta_{jx}$, $\beta_{jz}$, $\vartheta_{jy }$, $\vartheta_{jz}$), the linear scale factors ($k_{j1}$), and the angular motion ($T_1 \sim T_6$, $S_1 \sim S_6$, etc.). Because the magnitudes of the misalignment angles are of the order of $10^{-4}$, the more misalignment angles in an error term, the smaller its magnitude. Therefore, we analyze only those error terms with fewer than three misalignment angles. Coupling error terms with the same number of misalignment angles have the same order of magnitude. The basic coupling error terms with one misalignment angle are typically $\beta_{j{n}}k_{j1}S_{p}R_j$, $\vartheta_{j{m}}k_{j1}S_{p}R_j$, $\beta_{j{n}}k_{j1}T_{p}R_j$, and $\vartheta_{j{m}}k_{j1}T_{p}R_j$; the basic coupling error terms with two misalignment angles are typically $\beta_{j{n}}\vartheta_{j{m}}k_{j1}T_{p}R_j$ and $\beta_{j{n_1}}\beta_{j{n_2}}k_{j1}T_{p}R_j$; the basic coupling error term with three misalignment angles is typically $\beta_{j{n_1}}\beta_{j{n_2}}\vartheta_{m}k_{j1}T_{p}R_j$. In the RAGG analytical model, $D^{*}_{\beta_{jn}k_{j1}R_j}T_p$, $\sum_{\beta_{jn}k_{j1}R_j}S_p$, $D^{*}_{\beta_{j{n_1}}\beta_{j{n_2}}k_{j1}R_j}S_p$, $\sum_{\beta_{j{n_1}}\beta_{j{n_1}}k_{j1}R_j}T_p$, $\sum_{\beta_{j{n_1}}\theta_{j{m}}R_j}T_p$, $D^{*}_{\beta_{j{n_1}}\beta_{j{n_2}}\theta_{j{m}}R_j}T_p$, etc. are the error terms concerning the linear scale factors, misalignment angles, and angular motion. Here, we take  $\beta_{j{n}}k_{j1}T_{p}R_j$, $\beta_{j{n_1}}\beta_{j{n_2}}k_{j1}T_{p}R_j$, and $\beta_{j{n_1}}\beta_{j{n_2}}\vartheta_{m}k_{j1}T_{p}R_j$ as examples of analyzing the coupling error terms with one, two, and three  misalignment angles, respectively. Let $M_{e51}$, $M_{e52}$, and $M_{e53}$ represent the measurement errors contributed by the coupling error terms with one, two, and three misalignment angles, respectively:
 \begin{equation}\label{eq39}
 \begin{array}{l}
{M_{e51}} = {{{\beta _{jn}}{k_{j1}}{T_p}{R_j}} \mathord{\left/
 {\vphantom {{{\beta _{jn}}{k_{j1}}{T_p}{R_j}} {{k_{ggi}}}}} \right.
 \kern-\nulldelimiterspace} {{k_{ggi}}}},\\
{M_{e52}} = {{{\beta _{j{n_1}}}{\beta _{j{n_2}}}{k_{j1}}{T_p}{R_j}} \mathord{\left/
 {\vphantom {{{\beta _{j{n_1}}}{\beta _{j{n_2}}}{k_{j1}}{T_p}{R_j}} {{k_{ggi}}}}} \right.
 \kern-\nulldelimiterspace} {{k_{ggi}}}},\\
{M_{e53}} = {{{\beta _{j{n_1}}}{\beta _{j{n_2}}}{\vartheta _{jm}}{k_{j1}}{T_p}{R_j}} \mathord{\left/
 {\vphantom {{{\beta _{j{n_1}}}{\beta _{j{n_2}}}{\vartheta _{jm}}{k_{j1}}{T_p}{R_j}} {{k_{ggi}}}}} \right.
 \kern-\nulldelimiterspace} {{k_{ggi}}}}  \thickspace ,
\end{array}
 \end{equation}
where $k_{ggi}$ is the RAGG scale factor. Substituting ${{k}_{ggi}}=\sum\limits_{j=1}^{4}{{{k}_{j1}}R_j}$ into  Eq.~(\ref{eq39}), we obtain
 \begin{equation}\label{eq40}
 \begin{array}{l}
{M_{e51}} \approx 0.25{\beta _{jn}}{T_p},\\
{M_{e52}} \approx 0.25{\beta _{j{n_1}}}{\beta _{j{n_2}}}{T_p},\\
{M_{e53}} \approx 0.25{\beta _{j{n_1}}}{\beta _{j{n_2}}}{\vartheta _{jm}}{T_p}  \thickspace .
\end{array}
 \end{equation}
Let $T_{1critical}$, $T_{2critical}$, and $T_{3critical}$ represent the critical values for neglecting coupling error terms with one, two, and three misalignment angles, respectively. Similarly, assuming that $M_{e51}=M_{e52}=M_{e53}=0.1$~Eo and a misalignment angle of $\beta_{jn}=\vartheta_{jm}=10^{-4}$~rad, we calculate the critical values as $T_{1critical}=4 \times 10^3$~Eo, $T_{2critical}=4 \times 10^7$~Eo, and $T_{3critical}=4\times10^{11}$~Eo.
\begin{table}[htbp]
\centering
\caption{Conditions for Neglecting Error Terms Concerning Linear Scale Factors, Misalignment Angles, and Angular Motion}
\label{tab3}       
\begin{tabular}{ll}
\hline \hline
Error terms & Conditions\\ \hline
$\beta_{j{n}}k_{j1}T_{p}R_j$&
$\begin{array}{l}
\omega_{im}                   \le {\omega _{critical}} = {{{\rm{0}}{\rm{.057}}{{\rm{4}}^                   \circ }} \mathord{\left/
 {\vphantom {{{\rm{0}}{\rm{.057}}{{\rm{4}}^                   \circ }} s}} \right.
 \kern-\nulldelimiterspace} s}\\
{{\dot \omega }_{im}}                   \le {{{\dot \omega }}_{critial}} = {\rm{1}}{\rm{.71}}                  \times                   10^{-4}{{^                   \circ } \mathord{\left/
 {\vphantom {{^                   \circ } {{s^2}}}} \right.
 \kern-\nulldelimiterspace} {{s^2}}}
\end{array}$\\
$\beta_{j{n_1}}\beta_{j{n_2}}k_{j1}T_{p}R_j$&
$\begin{array}{l}
\omega_{im}                   \le {\omega _{critical}} = {{{\rm{5}}{\rm{.73}}{{\rm{4}}^                   \circ }} \mathord{\left/
 {\vphantom {{{\rm{5}}{\rm{.73}}{{\rm{4}}^                   \circ }} s}} \right.
 \kern-\nulldelimiterspace} s}\\
{{\dot \omega }_{im}}                   \le {{{\dot \omega }}_{critial}} = {\rm{1}}{\rm{.72}}{{^                   \circ } \mathord{\left/
 {\vphantom {{^                   \circ } {{s^2}}}} \right.
 \kern-\nulldelimiterspace} {{s^2}}}\\
\end{array}$\\
$\beta _{j{n_1}}\beta_{j{n_2}}\vartheta_{m}k_{j1}T_{p}R_j$ &
$\begin{array}{l}
\omega_{im}                   \le {\omega _{critical}} = {{{\rm{5}}{\rm{72}}{{\rm{4}}^                   \circ }} \mathord{\left/
 {\vphantom {{{\rm{5}}{\rm{.73}}{{\rm{4}}^                   \circ }} s}} \right.
 \kern-\nulldelimiterspace} s}\\
{{\dot \omega }_{im}}                   \le {{{\dot \omega }}_{critial}} = {\rm{1}}{\rm{.72}}                  \times                   10^4{{^                   \circ } \mathord{\left/
 {\vphantom {{^                   \circ } {{s^2}}}} \right.
 \kern-\nulldelimiterspace} {{s^2}}}\\
\end{array}$\\
\hline \hline
\end{tabular}
\end{table}
\lineskiplimit=0pt \lineskip=1pt
Correspondingly, we have calculated the conditions for neglecting the coupling error terms and listed them in Table~\ref{tab3}. Clearly, coupling error terms with more than one misalignment angle can be neglected.

\item Coupling error terms concerning linear scale factors, misalignment angles, and linear motion. The coupling error terms concerning linear scale factors, misalignment angles, and linear motion are  permutations of the misalignment angles ($\beta_{jx}$, $\beta_{jz}$, $\vartheta_{jy }$, $\vartheta_{jz}$), the linear scale factors ($k_{j1}$), and the linear motion ($a_x$, $a_y$, $a_z$). Similarly, we only analyze those coupling error terms that have fewer than three misalignment angles. Because coupling error terms that have the same number of misalignment angles also have the same magnitude, we take $\beta_{jn}k_{j1}a_{z}$, $\beta_{jn1}\beta_{jn2}k_{j1}a_{x}$, and $\beta_{jn1}\beta_{jn2}\vartheta_{jm}k_{j1}a_{x}$ as  examples of analyzing coupling error terms with one, two, and three misalignment angles, respectively. Let $M_{e61}$, $M_{e62}$, and $M_{e63}$ represent the measurement errors contributed by error terms with one, two, and three misalignment angles, respectively:
\begin{equation}\label{eq41}
\begin{array}{l}
{M_{e61}} = {{{\beta _{jn}}{k_{j1}}{a_z}} \mathord{\left/
 {\vphantom {{{\beta _{jn}}{k_{j1}}{a_z}} {{k_{ggi}}}}} \right.
 \kern-\nulldelimiterspace} {{k_{ggi}}}} \approx 0.25{{{\beta _{jn}}{a_z}} \mathord{\left/
 {\vphantom {{{\beta _{jn}}{a_z}} R}} \right.
 \kern-\nulldelimiterspace} R},\\
{M_{e62}} = {{{\beta _{j{n_1}}}{\beta _{j{n_2}}}{k_{j1}}{a_z}} \mathord{\left/
 {\vphantom {{{\beta _{j{n_1}}}{\beta _{j{n_2}}}{k_{j1}}{a_z}} {{k_{ggi}}}}} \right.
 \kern-\nulldelimiterspace} {{k_{ggi}}}} \approx 0.25{{{\beta _{j{n_1}}}{\beta _{j{n_2}}}{a_z}} \mathord{\left/
 {\vphantom {{{\beta _{j{n_1}}}{\beta _{j{n_2}}}{a_z}} R}} \right.
 \kern-\nulldelimiterspace} R},\\
{M_{e63}} = {{{\beta _{j{n_1}}}{\beta _{j{n_2}}}{\vartheta _{jm}}{k_{j1}}{a_z}} \mathord{\left/
 {\vphantom {{{\beta _{j{n_1}}}{\beta _{j{n_2}}}{\vartheta _{jm}}{k_{j1}}{a_z}} {{k_{ggi}}}}} \right.
 \kern-\nulldelimiterspace} {{k_{ggi}}}} \approx 0.25{{{\beta _{j{n_1}}}{\beta _{j{n_2}}}{\vartheta _{jm}}{a_z}} \mathord{\left/
 {\vphantom {{{\beta _{j{n_1}}}{\beta _{j{n_2}}}{\vartheta _{jm}}{a_z}} R}} \right.
 \kern-\nulldelimiterspace} R}   \thickspace .
\end{array}
\end{equation}
Let $a_{1critical}$, $a_{2critical}$, and $a_{3critical}$ represent the critical values for neglecting coupling error terms with one, two, and three misalignment angles, respectively. Similarly, assuming that $M_{e61}=M_{e62}=M_{e63}=0.1$~Eo, $R=0.1$~m, and a misalignment angle of $\beta_{jn}=\vartheta_{jm}=10^{-4}$~rad, the critical values are calculated as $a_{1critical}=4                  \times                   10^{-8}$~g, $a_{2critical}=4                  \times                   10^{-4}$~g, and $a_{3critical} = 4$~g, respectively. Because in gravity gradiometry the RAGG acceleration is usually in the order of 0.1~g, we can neglect coupling error terms with more than two misalignment angles.
\end{compactenum}
\subsubsection{The simplified RAGG analytical model}
Based on the previous analysis and neglecting the small coupling error terms that have little effect on the RAGG sensitivity, we obtain the following simplified RAGG analytical model:
\begin{equation}\label{eq42}
\begin{array}{l}
{G^{s}_{out}} =A_{2\Omega }^s\sin 2\Omega t + A_{2\Omega }^c\cos 2\Omega t + A_\Omega ^s\sin \Omega t\\[2pt]
 + A_\Omega ^c\cos \Omega t + {A_0} \thickspace ,
\end{array}\end{equation}
where $A_{2\Omega }^s$ and $A_{2\Omega }^c$ are the amplitudes of $\sin2\Omega t$ and $\cos2\Omega t$, respectively, namely
\begin{equation}\label{eq43}
\begin{array}{l}
A_{2\Omega }^s = {T_1}\sum {_{{k_1}R}}  + {T_2}\sum {_{({\vartheta _Z} - 2{\beta _Z}){k_1}R}} \\ [3.5pt] + {a_x}{a_y}\sum {_{{k_5} - {k_2}}}
  - 0.5\sum {_{{k_6}}} \left( {{a_x}^2 - {a_y}^2} \right)\\[3.5pt]
A_{2\Omega }^c =  - {T_1}\sum {_{({\vartheta _Z} - 2{\beta _Z} ){k_1}R}}  + {T_2}\sum {_{{k_1}R}} \\[3.5pt]
 + 0.5\left( {{a_x}^2 - {a_y}^2} \right)\sum {_{{k_5} - {k_2}}}  + {a_x}{a_y}\sum {_{{k_6}}}
\end{array}
\end{equation}
In Eq.~(\ref{eq42}), $A_{\Omega }^s$ and $A_{\Omega }^c$ are the amplitudes of  $\sin\Omega t$ and $\cos\Omega t$, respectively, namely
\begin{align}\label{eq45}
A_\Omega ^s{\rm{ }} &= {\rm{  }}D_{ - (1 + {\beta _z}{\vartheta _z}){k_1},({\vartheta _Z} - {\beta _Z} + {\beta _X}{\vartheta _Y}){k_1}}^{12,34}{a_x}
\notag \\[3.5pt] &
 + {\rm{ }}D_{({\vartheta _z} - {\beta _z} + {\beta _X}{\vartheta _Y}){k_1},(1 + {\beta _z}{\vartheta _z}){k_1}}^{12,34}{a_y} + {\rm{ }}D_{{k_4}, - {k_8}}^{12,34}{a_x}{a_z}
\notag \\[3.5pt] &
 - D_{{k_8},{k_4}}^{12,34}{a_y}{a_z} - D_{{\theta _Y}R}^{12}{T_3}{\rm{ }} - D_{{\theta _Y}R}^{34}{T_4}
\notag \\[3.5pt] &
 + D_{{\beta _X}{k_1}R}^{12}{\rm{ }}{S_1} - {\rm{ }}D_{{\beta _X}{k_1}R}^{34}{S_2}{\rm{ }} \notag \\[3.5pt]
A_\Omega ^c & = D_{({\vartheta _z} - {\beta _z} + {\beta _X}{\vartheta _Y}){k_1},(1 + {\beta _z}{\vartheta _z}){k_1}}^{12,34}{a_x}
\notag \\[3.5pt] &
 + D_{(1 + {\beta _z}{\vartheta _z}){k_1}, - ({\vartheta _z} - {\beta _z} + {\beta _X}{\vartheta _Y}){k_1}}^{12,34}{a_y} - D_{{k_8},{k_4}}^{12,34}{a_x}{a_z}
 \notag \\[3.5pt] &
 + {\rm{ }}D_{ - {k_4},{k_8}}^{12,34}{a_y}{a_z} - {\rm{ }}D_{{\theta _Y}R}^{12}{T_4}{\rm{  }} + D_{{\theta _Y}R}^{34}{T_3}
 \notag \\[3.5pt] &
 - {\rm{ }}D_{{\beta _X}{k_1}R}^{34}{S_1} - D_{{\beta _X}{k_1}R}^{12}{\rm{ }}{S_2}
\end{align}
In Eq.~(\ref{eq42}), $A_0$ is given by
\begin{align}\label{eq47}
{A_0} &= 0.5D_{{k_2} + {k_5}}^{1234}({a_x}^2 + {a_y}^2) + D_{{k_7}}^{1234}{a_z}^2 \notag\\[3.5pt]&
+ D_{{\theta _Y} - {\beta _X}{\theta _Z}}^{1234}{a_z} - D_{{\theta _Z}R}^{1234}{T_5} + D_{{k_1}R}^{1234}{T_6} + D_{{k_0}}^{1234}
\end{align}
In the simplified analytical model, the error propagation coefficients in $\sin2\Omega t$ and $\cos2\Omega t$ are of the form $\sum {_*}$; $\left( {\sum {_{{k_5-k_2}}}} \right)$ and $\sum {_{{k_6}}}$ directly make the DC component of the linear acceleration into the output of the RAGG. The error propagation coefficients in $\sin\Omega t$ and $\cos\Omega t$ are in the forms of $D^{12}_{*}$, $D^{34}_{*}$;
$D^{12}_{*}$ and $D^{34}_{*}$ are caused by mismatch of one pair of accelerometer parameters, such as scale factors or mounting errors.
The error propagation coefficients in $A_0$ are of the form $D^{1234}_{*}$; similarly, $D^{1234}_{*}$ are caused by the mismatch of two pairs of accelerometer parameters. By these error propagation coefficients, the low-frequency components of the linear motion and angular motion are transferred into outputs of the RAGG. The angular motion and vibration are isolated by using a stabilized platform. In the process of moving-base gravity gradiometry, the linear acceleration has more impact on the RAGG than does the angular motion. From Eqs.~(\ref{eq42}) $\sim$ (\ref{eq47}), we can apply the linear acceleration of a specific frequency to the RAGG, producing a detection signal for on-line continuous compensation of the error propagation coefficients. We can also calibrate the error propagation coefficients and record the line motion and angular motion of the RAGG in gravity gradiometry for off-line error compensation. However, because compensating for error propagation coefficients is not the focus of the present work, we do not discuss it in depth.

\subsection{RAGG Numerical Model}\label{sec3.3}
\begin{figure}[!ht]\centering
\subfloat[Principle of the RAGG numerical model.]{\includegraphics[width=0.4\textwidth]{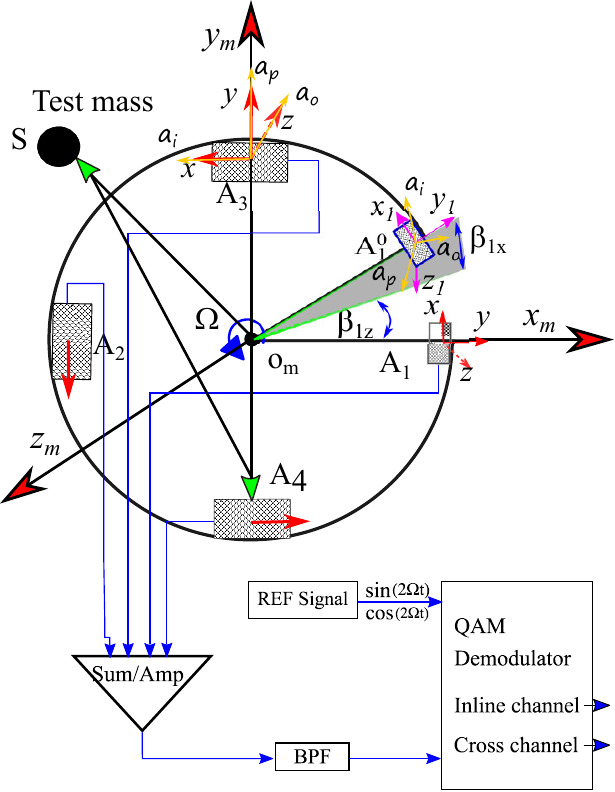}\label{fig3a}}\\
\subfloat[Program flow of the RAGG numerical model.]{\includegraphics[width=0.4\textwidth]{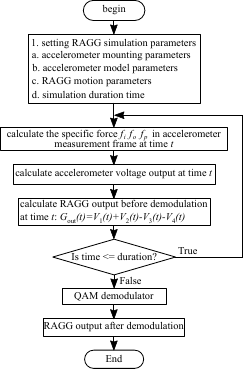}\label{fig3b}}
	\caption{Principle and program flow of the RAGG numerical model.}\label{fig3}
\end{figure}

We have established a high-precision numerical model of the RAGG, as shown in Fig.\ref{fig3}.
In the numerical model, each accelerometer has six mounting error parameters: radial distance ($R_j$), initial phase angle ($\beta_{jz}$), altitude angle ($\beta_{jx}$), and misalignment error angles ($\vartheta_{jx}$, $\vartheta_{jy}$, and $\vartheta_{jz}$). Among them, the radial distance ($R_j$), initial phase angle ($\beta_{jz}$), and altitude angle ($\beta_{jx}$) determine the mounting position of the accelerometer; the misalignment error angles ($\theta_{jx}$, $\theta_{jy}$ and $\theta_{jz}$) determine
the orientation deviation between the accelerometer measurement frame and the accelerometer nominal frame of the actual
mounting position.
Moreover, each accelerometer has nine other output model parameters: zero bias ($K_{j0}$), linear scale factors ($K_{j1}$), second-order error coefficients ($K_{j2}$, $K_{j4}$, $K_{j5}$, $K_{j6}$, $K_{j7}$, $K_{j8}$), and
 current to voltage gain ($k_{jV/I}$). In total, each accelerometer has 15 parameters. We use a test mass to produce gravitational gradients to excite the RAGG. The specific force of the accelerometer ${{A}_{j}}$ in the RAGG numerical model is given by:
\begin{equation}\label{eq49}
\begin{array}{*{20}{l}}
{{{\bm{f}}_j} = {{\bm{f}}_{cmm}} + {{{\bm{\dot \omega }}}_{im}}                   \times                   {{\bm{r}}_{{o_m}{A_j}}} + {{\bm{\omega }}_{im}}                   \times                   \left( {{{\bm{\omega }}_{im}}                   \times                   {{\bm{r}}_{{o_m}{A_j}}}} \right)}\\
{ - {{Gm{{\bm{A}}_{\bm{j}}}{\bm{S}}} \mathord{\left/
 {\vphantom {{Gm{{\bm{A}}_{\bm{j}}}{\bm{S}}} {{{\left| {{{\bm{A}}_{\bm{j}}}{\bm{S}}} \right|}^3}}}} \right.
 \kern-\nulldelimiterspace} {{{\left| {{{\bm{A}}_{\bm{j}}}{\bm{S}}} \right|}^3}}}{\rm{ }}} \thickspace.
\end{array}
\end{equation}
Where $\bm{f}_{cmm}$ is the specific force of the RAGG; ${{\dot {\bm{\omega }} }_{im}}$ is the angular acceleration of the RAGG with respect to the inertial frame; ${\bm{\omega } _{im}}$ is the angular velocity of the RAGG with respect to the inertial frame; $G$ is the gravitational constant; ${{{\bm{r}}_{{o_m}{A_j}}}}$ is the position vector of accelerometer $A_j$ in the RAGG measurement; $m$ represents the weight of the test mass; and
${{{\bm{A}}_{\bm{j}}}{\bm{S}}}$ is the position vector from  accelerometer $A_j$ to the test mass. If the test mass is not a point mass, the gravitational acceleration that the RAGG accelerometers undergo produced by the test mass can be calculated using finite element analysis.
In addition, the test mass can be in motion with respect to the RAGG; in this case, $\bm{A_jS}$ is time varying \cite{yu2018calibration}.
 The specific forces of  accelerometer $A_j$ in the accelerometer nominal frame of the actual mounting position ($f_{jx}$, $f_{jy}$, $f_{jz}$) can be calculated from:
  \begin{equation}\label{eq50}
{\begin{array}{*{20}{l}}
  {{f_{jx}} = {{\bm{f}}_j}                   \cdot {{\bm{\tau }}_{jx}}} \thickspace,\\
  {{f_{jy}} = {{\bm{f}}_j}                   \cdot {{\bm{\tau }}_{jy}}} \thickspace,\\
  {{f_{jz}} = {{\bm{f}}_j}                   \cdot {{\bm{\tau }}_{jz}}}\thickspace.
\end{array}}
\end{equation}
Where $\bm{\tau }_{jx}$, $\bm{\tau }_{jy}$, and $\bm{\tau }_{jz}$ are unit vectors of the accelerometer nominal frame of the actual mounting position in the directions of the x-, y-, and z-axes. The specific forces in the accelerometer measurement frame are:
\begin{equation}\label{eq51}	
\left[ \begin{matrix}
   {{f}_{ji}}  \\
   {{f}_{jo}}  \\
   {{f}_{jp}}  \\
\end{matrix} \right]=\bm{C}\left[ \begin{matrix}
   {{f}_{jx}}  \\
   {{f}_{jy}}  \\
   {{f}_{jz}}  \\
\end{matrix} \right]\thickspace,
\end{equation}	
  where $\bm{C}$ is the transformation matrix from the accelerometer nominal frame of the actual mounting position to the
accelerometer measurement frame; $\bm{C}$ is given in Eq.\eqref{eq2}.
 To make the numerical model approximate the actual RAGG, we add accelerometer noise to the accelerometer model:
\begin{equation}\label{eq52}
\begin{array}{l}
\frac{{{V_j}}}{{{k_{j{V \mathord{\left/
 {\vphantom {V I}} \right.
 \kern-\nulldelimiterspace} I}}}{K_{j1}}}} = {f_{jnoise}} + {f_{ji}} + {K_{j0}} + {K_{j2}}{f_{ji}}^2 + {K_{j5}}{f_{jo}}^2
\\[2pt]
 + {K_{j7}}{f_{jp}}^2 + {K_{j6}}{f_{ji}}{f_{jo}}
 + {K_{j4}}{f_{ji}}{f_{jp}} + {K_{j8}}{f_{jo}}{f_{jp}}\thickspace.
\end{array}
\end{equation}
The accelerometer noise $f_{jnoise}$ is simulated by a power spectral density model:
\begin{equation}
\Phi (f)_{noise}  = \alpha {f^{ - b}} + {\omega _T} \thickspace,
\end{equation}
where $\alpha$ and $b$ represent the amplitude and low-frequency growth of the red noise, and ${\omega _T}$ denotes the amplitude of the white noise.

Fig.\ref{fig3b} is the program flow of the RAGG numerical model. Firstly, the RAGG simulation parameters are set up, including test masses parameters, RAGG rotating disk parameters, accelerometer mounting parameters, accelerometer model parameters, RAGG motion parameters, etc.
Then substituting the parameters into the formula ~\eqref{eq49} $\sim$ \eqref{eq51} calculates the specific force in accelerometer measurement frame at time $t$. According to the formula ~\eqref{eq52}, calculating the output voltage of the RAGG accelerometer,
the RAGG output before demodulation at time t is calculated by: $G_{out}(t)=V_1(t)+V_2(t)-V_3(t)-V_4(t)$.
The above process is repeated until time $t$ is equal to the simulation duration time.
Finally, the RAGG output data is input to the quadrature amplitude modulation (QAM) demodulator to extract gravitational gradient.

 Let $\Gamma_{xx}$, $\Gamma_{xy}$, $\Gamma_{xz}$, $\Gamma_{yy}$, and $\Gamma_{yz}$ represent the five independent gravitational gradient elements at the origin of the RAGG measurement frame. When mass is far enough away from the RAGG, the gravitational acceleration measured by the RAGG accelerometers is a first-order approximation of the gravitational acceleration and gravitational gradient tensor at the center of the rotating disc; in this case, the inline channel measurement and the cross channel measurement
of the RAGG approximate $\Gamma_{xx}-\Gamma_{xy}$ and $\Gamma_{xy}$; otherwise, the inline channel measurement and cross channel measurement of the RAGG are the sum of $\Gamma_{xx}-\Gamma_{xy}$, $\Gamma_{xy}$, and high-order gravitational gradient tensor elements.
To distinguish between $\Gamma_{xx}-\Gamma_{xy}$, $\Gamma_{xy}$ and the measurements of the RAGG, we call $\Gamma_{xx}-\Gamma_{xy}$, $\Gamma_{xy}$ center gravitational gradients; $\Gamma_{xx}-\Gamma_{xy}$ is the inline channel of the center gravitational gradients;
$\Gamma_{xy}$ is the cross channel of the center gravitational gradients.
As mentioned preciously, in the analytical model, the gravitational acceleration that the RAGG accelerometer
undergo is a first-order approximation of the gravitational acceleration and gravitational gradient tensor at the center of the rotating disc; but in the numerical model, the gravitational accelerations are calculated using Newton’s law of gravitation instead of a linear approximation; therefore, the outputs of the analytical model are close to the center gravitational gradients; and those of the numerical model are close to the actual measurements of the RAGG.

\renewcommand{\multirowsetup}{\centering}

\begin{table}[htbp]
\centering
  \caption{Error Sources and Characteristics of the Three RAGG Models}\label{tab4}
    \begin{tabularx}{85mm}{lX} \hline \hline
    Model  categories & Model characteristics\\ \noalign{\smallskip} \hline
    \noalign{\smallskip}
    Numerical model & It has considered almost all of the error sources,
    such as  accelerometer mounting position error, accelerometer input axis misalignment,
     circuit gain mismatch, accelerometer linear scale factor imbalance, accelerometer
      second-order error coefficients, high order gravitational gradient tensor. It approaches the real RAGG. \\
      \noalign{\smallskip}
    Analytic model & Compared with the numerical model,
    it principally neglects high order gravitational gradient tensor, the coupling error propagation coefficients  concerning
     accelerometer second-order error coefficients and misalignment angles. \\
     \noalign{\smallskip}
    \mrowtabcel{l}{Simplified \\ analytical  model}  & Compared with the analytic model,
    it principally neglects high order gravitational gradient tensor, the coupling terms concerning accelerometer second-order error coefficients, angular motions, and linear motions,
    the coupling terms of accelerometer second-order error coefficients and angular motions.     \\ \hline \hline
    \end{tabularx}%
\end{table}%
In summary, based on the principle and configuration of the RAGG, we established three RAGG models, namely a numerical model, an analytical model, and a simplified analytical model. The numerical model considers most of the error sources and can simulate the actual RAGG. The analytical model is an approximate model of the RAGG (or the numerical model); it neglects the high order gravitational gradient tensor, the coupling error propagation coefficients concerning the accelerometer second-order error coefficients and misalignment angles. The simplified analytical model is a simplified version of the analytical model. The error sources and model characteristics of these three models are listed in Table~\ref{tab4}.

\section{Experiment}
In this section, a multi-frequency gravitational gradient simulation experiment and a dynamic simulation experiment are designed to verify the correctness of the three RAGG models; the turbulence simulation experiments are designed to evaluate the performance of analytical models.

In multi-frequency gravitational gradient simulation experiment, the three RAGG models simulate a measurement scene, in which a test mass is rotated around a perfect RAGG for producing multi-frequency gravitational gradient exciting the RAGG. In this case, the theoretical measurements of a RAGG are the center gravitational gradients; the outputs of the analytical models, that of the numerical model, and the center gravitational gradients
can be consistent with each other, only when the analytical models and the numerical model don't have principle errors and calculation errors.
Besides based on the properties of three RAGG models, the outputs of the analytical models are more closer to the center gravitational gradients than that of the numerical model. Compared the multi-frequency gravitational gradient experimental results to the theoretical ones, we can preliminarily verify the correctness of the three RAGG models.

Multi-frequency gravitational gradient simulation experiment only can verify the correctness of the three RAGG models when a RAGG doesn't have imperfect factors; furthermore dynamic simulation experiment is designed to verify the correctness of the three RAGG models when a RAGG has imperfect factors. In the dynamic simulation experiment, the three RAGG models simulate a measurement scene, in which a imperfect RAGG with accelerometer mounting errors, accelerometer second order error coefficients, accelerometer linear scale factors imbalance, etc.
undergoes linear motion and angular motion. Based on the properties of three RAGG models, if there are no principle errors and calculation errors in RAGG numerical model and RAGG analytical models, then the outputs of the RAGG analytical models, that of the numerical model should be consistent with each other, and the outputs of analytical model should be more closer to the numerical model than that of the simplified analytical model.
Compared the dynamic experimental results to the theoretical ones, we can verify the correctness of the three RAGG models.

In turbulence simulation experiments, the three RAGG models simulate a imperfect RAGG encountering air turbulence;
RAGG numerical model is used as a real RAGG to evaluate the noise floor of the analytical models, which is limits of accuracy in error compensation method based on the analytical models.

\subsection{Multi-frequency gravitational gradient simulation experiment}
In multi-frequency gravitational gradient simulation experiment, a test mass rotates about the RAGG with time-varying angular velocity producing multi-frequency gravitational gradient excitations. Based on the angular velocity of the test mass and its initial coordinate in the RAGG measurement frame, we can obtain the coordinates of the test mass in the RAGG measurement frame at any time. We then calculate the gravitational gradient tensor at the origin of the RAGG measurement frame and then calculate the center gravitational gradients.
\begin{figure}[!t]\centering
	\includegraphics[width=0.4\textwidth]{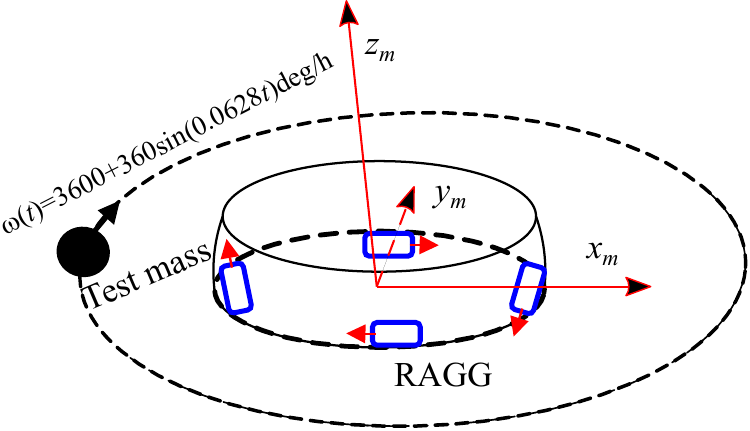}
	\caption{ A test mass rotating about the RAGG with time-varying angular velocity.}\label{fig4}
\end{figure}
The three RAGG models simulate a perfect RAGG with no accelerometer mounting errors, accelerometer scale-factor imbalances, or accelerometer second-order error coefficients, so we set the accelerometer mounting errors ($dR_j$, $\beta_{jx}$, $\beta_{jz}$, $\vartheta_{jx}$, $\vartheta_{jy}$, $\vartheta_{jz}$) and the accelerometer second-order error coefficients ($k_{j2}$, $k_{j4}$, $k_{j5}$, $k_{j6}$, $k_{j7}$, $k_{j8}$) to zero. The linear scale factor of the four accelerometers is $k_{j1}=10$~mA/g, the current-to-voltage gain is $k_{jV/I}=10^9$~ohm, the nominal mounting radius $R$ is 0.1~m, and the rotation frequency of the RAGG disc is 0.25~Hz. Fig.~\ref{fig4} shows a point mass of 486~kg with an initial position in the RAGG measurement frame of ($1.2,0,0$) and rotating about the RAGG with time-varying angular speed $\omega (t)=3,600+360\sin(0.0628t)^\circ/{\rm{h}}$.
\begin{figure}[!t]\centering
	\includegraphics[width=0.4\textwidth]{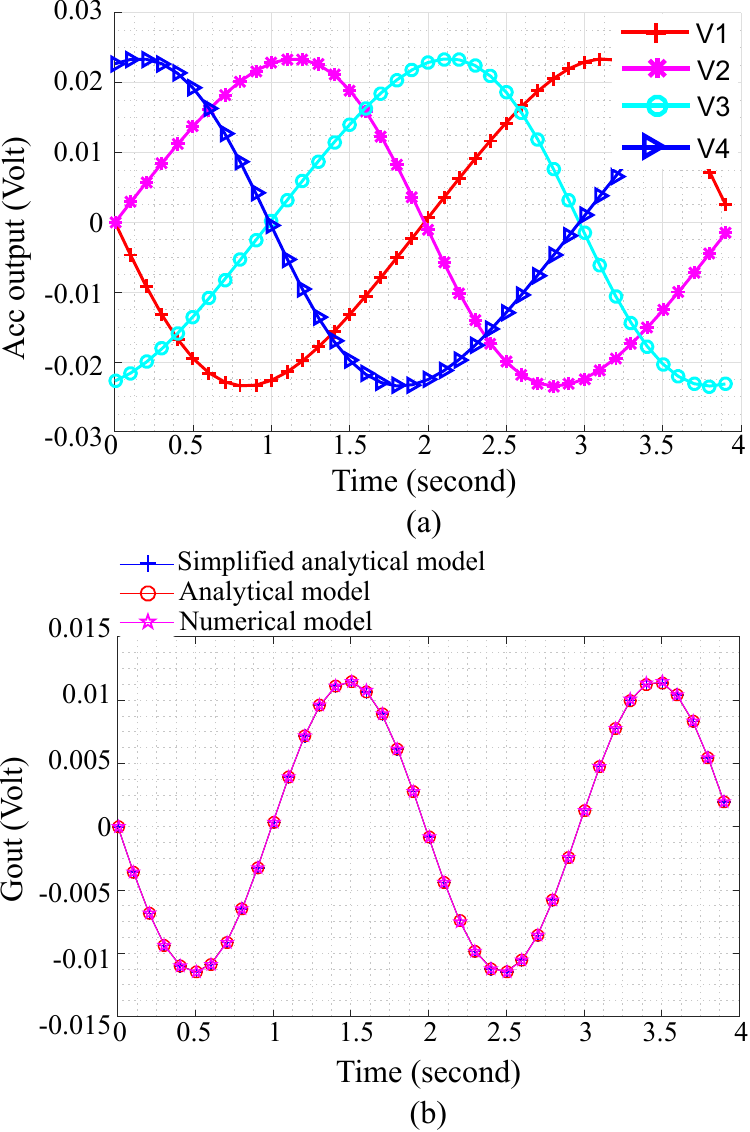}
	\caption{(a) Accelerometer voltage outputs in RAGG numerical model. (b) Output voltage before demodulation comparison among the numerical model, the analytical model, and the simplified analytical model.}\label{fig5}
\end{figure}
\begin{figure}[!t]\centering
	\includegraphics[width=0.4\textwidth]{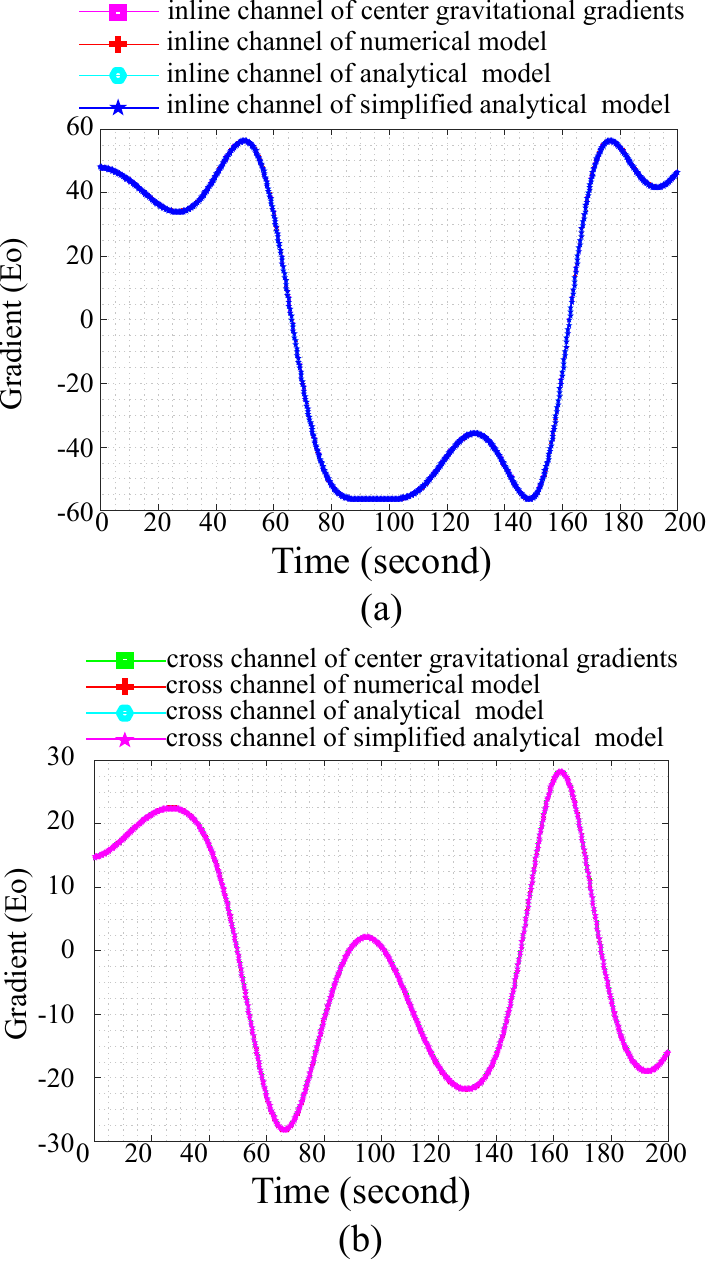}
	\caption{Demodulated gravitational gradient comparison among the three RAGG models and the center gravitational gradients. (a) Inline channel. (b) Cross-channel.}\label{fig6}
\end{figure}
\begin{figure}[!t]\centering
	\includegraphics[width=0.4\textwidth]{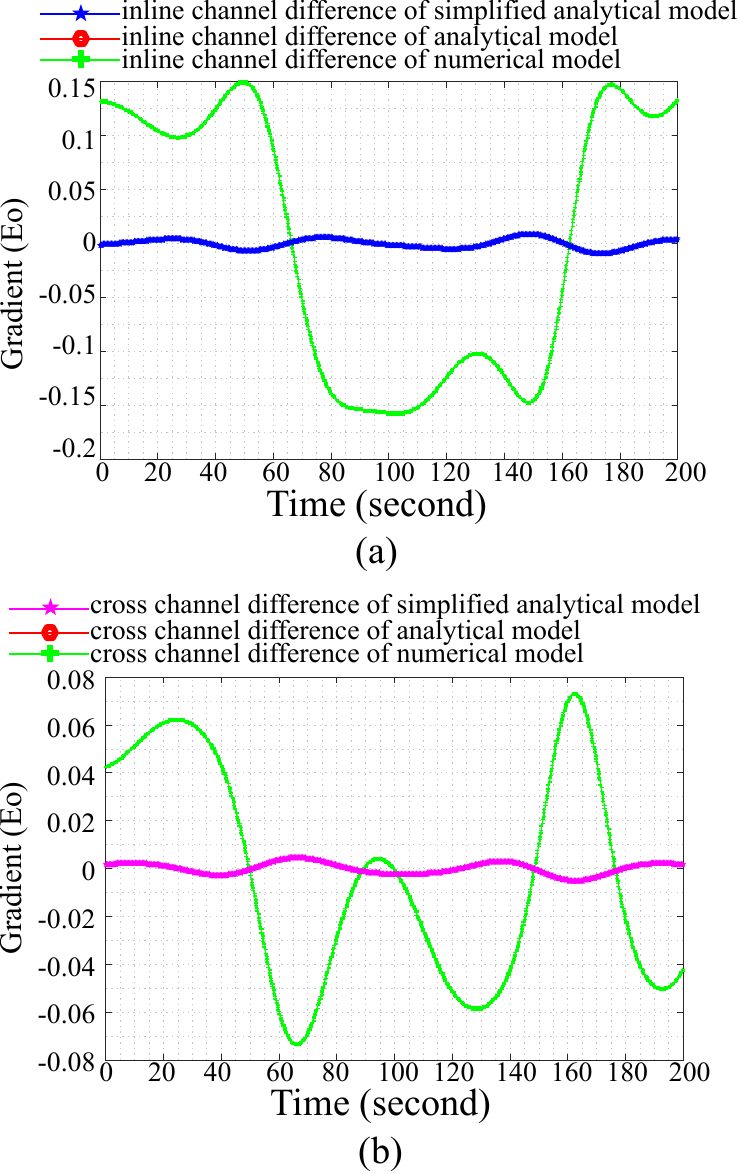}
	\caption{Difference among the three RAGG models and the center gravitational gradients. (a) Inline channel. (b) Cross-channel.}\label{fig7}
\end{figure}

Substituting the above experimental parameters into the analytical model \eqref{eq17} $\sim$ \eqref{eq19} and
the simplified analytical model \eqref{eq42} $\sim$ \eqref{eq47} calculate the output of the analytical model and that of the simplified analytical model
before demodulation. Inputting these experimental parameters into the numerical model,
according to the program flow shown in Fig.\ref{fig3b}, calculate the output of the numerical model before demodulation.
Finally, the output of the analytical model, that of the simplified analytical model, that of the numerical model are demodulated by the same QAM demodulator.

Fig.~\ref{fig5}(a) shows the voltage outputs of the four accelerometers in the RAGG numerical model excited by the rotating point mass. Fig.~\ref{fig5}(b) shows the output voltage before demodulation of the numerical model, that of the analytical model, and that of the simplified analytical model; the output voltages before demodulation of the three RAGG models are consistent with each other. Fig.~\ref{fig6} shows the demodulated gravitational gradient comparison among the three RAGG models and the center gravitational gradients; from Fig.~\ref{fig6}(a) and Fig.~\ref{fig6}(b), we can see that the inline channel and the cross channel of the three RAGG models are consistent with those of the center gravitational gradients. Besides, we have calculate the differences between the center gravitational gradients and the outputs of the RAGG model. To simplify the description, the difference between outputs of the RAGG numerical model and the center gravitational gradients is called difference of the numerical model; similarly, the difference between outputs of the analytical model and the center gravitational gradients is called difference of the analytical model. Fig.~\ref{fig7} shows the difference of the numerical model, the difference of the analytical model, and the difference of the simplified analytical model.
From Fig.\ref{fig7}, the difference of the analytical model and the difference of the simplified analytical model
are consistent with each other, and in the order of $10^{-3}$~Eo, but that of the numerical model is in the order of $10^{-1}$~Eo. The difference of the analytical model and the difference of the simplified analytical model are much smaller than the difference of the numerical model. The reason is that the high-order gravitational gradient tensor causes the outputs of the numerical model or the measurements of the RAGG to deviate from the center gravitational gradients. These experimental phenomenons are consistent with the theoretical ones.

\subsection{Dynamic Simulation Experiment}
In dynamic simulation experiment, the accelerometer parameters of the three RAGG models are listed in Table~\ref{tab5}. The nominal mounting radius $R$ is 0.1~m and the rotation frequency of the disc is 0.25~Hz. The scale factor of the RAGG is $k_{ggi} = k_{11}R_1+k_{21}R_2+k_{31}R_3+k_{41}R_4 = 4.54               \times10^{-4}$~V/Eo, and the demodulation filtering in QAM demodulator is typically achieved with a FIR low-pass filter with a cut-off frequency at 0.2 Hz. In airborne gravity gradiometry, the linear accelerations of the RAGG are in the order of 0.1~g, and the harmonic components of angular motion and linear motion whose fundamental frequencies equal the rotation rate of the rotating disc have considerable impact on the RAGG sensitivity, in the dynamic experiment, the angular velocities and linear accelerations consists of DC components and first harmonic components. The motion parameters of the RAGG are listed in Table~\ref{tab6}.
\renewcommand{\multirowsetup}{\centering}
\begin{table}[]
\centering
\caption{Accelerometer Mounting Parameters and Output Model Parameters in the Three RAGG Models}
\label{tab5}
\begin{tabular}{lllll} \hline \hline
  \multirow{2}{*}{Parameter name (unit)}                          & \multicolumn{4}{l}{ \centering Accelerometer name } \\
                                                           & Acc.1    & Acc.2   & Acc.3   & Acc.4   \\ \hline
                        $dR_j$       ({$\mu$}m)       &0         &30       &35       &25     \\
                        $\beta_{jx}$ (arsec)            &21        &14      &17       &19      \\
                       $\beta_{jz}$ (arsec)             &18        &8       &10       &20       \\
                       $\vartheta {jy}$ (arsec)          &21        &10       &7       &19       \\
                       $\vartheta {jz}$ (arsec)          &30        &19       &20       &9        \\
                       $k_{j0}$ ({$\mu$}g)            &10        &2        &5        &8        \\
                       $k_{j1}$ (mA/g)                  &10        &11       &11.5     &12        \\
                        $k_{j2}$ ({$\mu$}g/g$^2$)     &5         &1        &8        &7        \\
                        $k_{j4}$ ({$\mu$}g/g$^2$)     &6         &12       &5        &16        \\
                        $k_{j5}$ ({$\mu$}g/g$^2$)     &4         &11       &9        &13        \\
                         $k_{j6}$ ({$\mu$}g/g$^2$)    &3         &12       &8        &10        \\
                        $k_{j7}$ ({$\mu$}g/g$^2$)     &5         &15       &7        &17        \\
                        $k_{j8}$ ({$\mu$}g/g$^2$)     &0         &7        &11       &20        \\
                         $k_{jV/I}$ (ohm)             &$10^9$    &$10^9$   &$10^9$   &$10^9$        \\   \hline \hline
\end{tabular}
\end{table}
\begin{table}[]
\centering
\caption{Linear Motion and Angular Motion Parameters in Dynamic Simulation Experiment}
\label{tab6}
\begin{tabular}{lllll} \hline \hline
               {Linear acceleration (g)}  &  Angular velocity (deg/h) \\ \hline
                 $a_x(t)=0.08+0.005 sin\Omega t$    & $\omega_{imx}(t)=1000+700 sin\Omega t $ \\
                 $a_y(t)=0.06+0.006 sin\Omega t$    & $\omega_{imy}(t)=1500+900 sin\Omega t $ \\
                 $a_z(t)=0.1+0.05 sin\Omega t$    & $\omega_{imz}(t)=2000+800 sin\Omega t $ \\  \hline \hline
\end{tabular}
\end{table}
\begin{figure}[!t]\centering
	\includegraphics[width=0.4\textwidth]{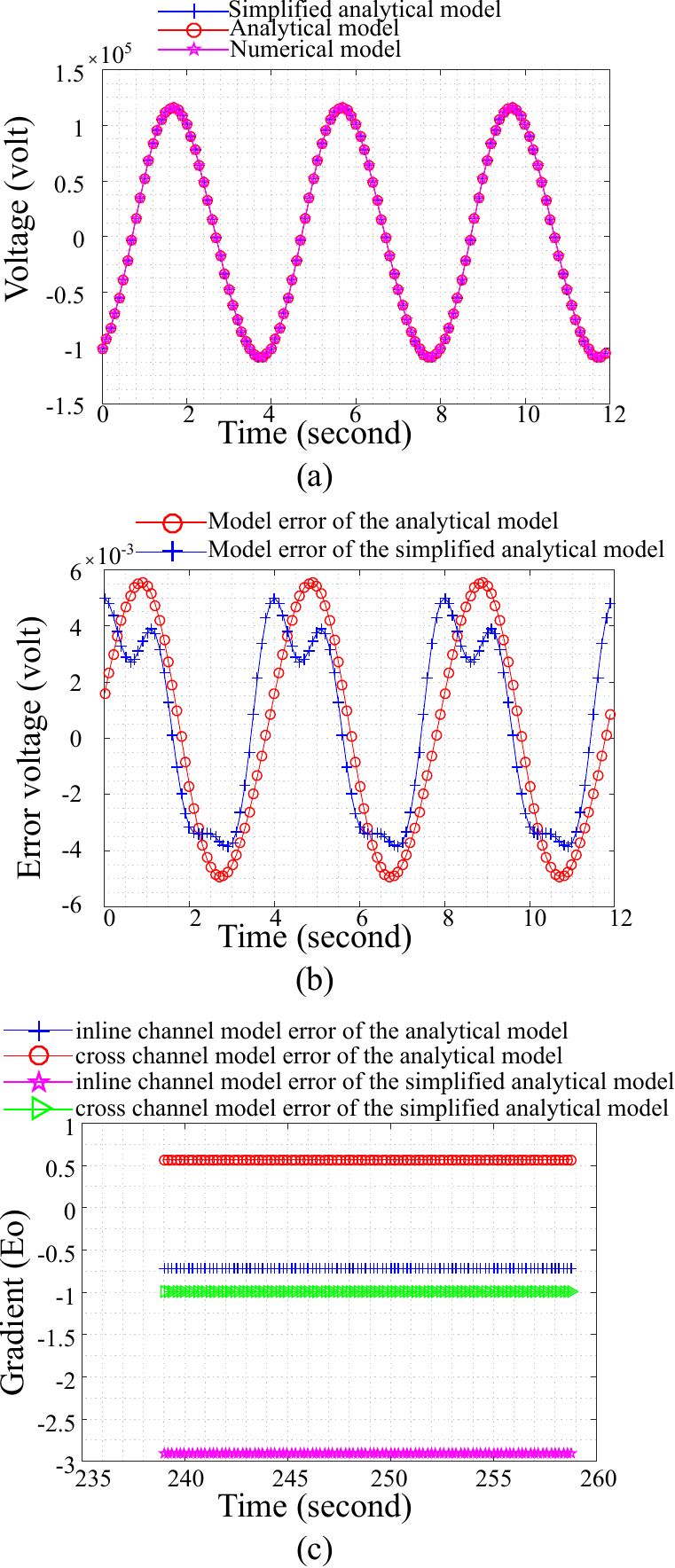}
	\caption{Results of dynamic simulation experiment. (a) Output voltage before demodulation comparison among the three RAGG models. (b) Model errors before demodulation. (c) Model errors after demodulation.}\label{fig8}
\end{figure}

The outputs of the analytical model and the simplified analytical model before demodulation are the combined signals of the four accelerometers. We treat the output before demodulation of the numerical model as that of the actual RAGG, and treat the differences between the analytical model and the numerical model as the model errors.
Fig.~\ref{fig8}(a) shows the outputs before demodulation of the numerical model, the analytical model, and the simplified analytical model. Fig.~\ref{fig8}(b) shows the model errors of the analytical model and the simplified analytical model before demodulation. Fig.~\ref{fig8}(c) shows the model errors of the analytical model and the simplified analytical model after demodulation. The inline-channel and cross-channel errors of the analytical model are roughly -~0.72~Eo, ~0.56~Eo, and those of the simplified analytical model are roughly -2.9~Eo, -0.99~Eo. From Fig.\ref{fig8}(a) $\sim$ Fig.\ref{fig8}(c), the outputs of the analytical model and the simplified analytical model are clearly consistent with that of the numerical model; the model errors of the analytical model are smaller than those of the simplified analytical model, thus the analytical model is more precise than the simplified analytical model. These experimental results are consistent with the theoretical ones. The output of the numerical model is of the order of $10^5$~V and the model errors of the analytical model and the simplified analytical model are of the order of $10^{-3}$~V. Therefore, the relative errors of the analytical model and the simplified analytical model are of the order of $10^{-8}$.

In Fig.~\ref{fig8}(a), the output voltage before demodulation is of the order of $10^5$~V, but the voltage contributed from the gravitational gradient is only of the order of $10^{-2}$~V. Most of the output voltages are caused by the joint effects of linear acceleration, angular acceleration, accelerometer mounting errors, accelerometer second-order error coefficients, and zero bias. In the dynamic experiment, although the misalignment angles are of the order of $10^{-4}$~rad and the accelerometer seconder-order error coefficients are of the order of $10^{-6}$~g/g$^2$, the RAGG is still very sensitive to its own motion. On-line error compensation is required to avoid tiny movements of the RAGG saturating or damaging the RAGG.

\subsection{Turbulence Simulation Experiment}
In gravity gradiometer data, the dynamic noise (motion error) caused by the air turbulence is a major source of error.
In gravity gradiometry, by recording the linear motion and angular motion of the RAGG, based on the analytical model or the simplified analytical model, we can calculate the motion error and compensate it; clearly, the limits of the error compensation accuracy equals the noise floor of the RAGG analytical model. In this section, we performed three turbulence experiments to evaluate noise floor of the analytical model and the simplified analytical model at air turbulence of different intensities.

\begin{figure}[!t]\centering
	\includegraphics[width=0.4\textwidth]{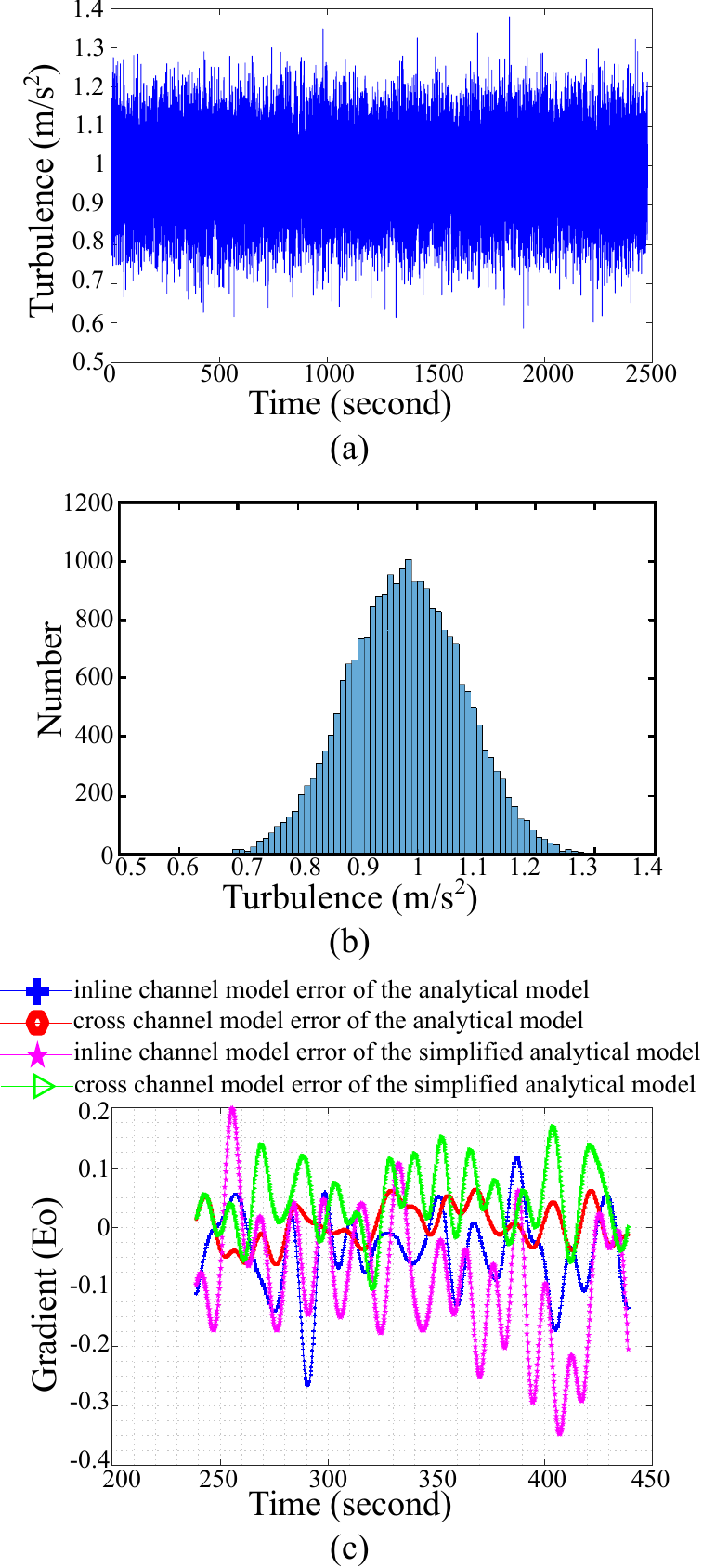}
	\caption{Results of the first turbulence simulation experiment. (a) Vertical acceleration caused by turbulence. (b) A histogram of vertical acceleration. (c) The analytical model error and the simplified analytical model error.}\label{fig9}
\end{figure}
\begin{figure}[!t]\centering
	\includegraphics[width=0.4\textwidth]{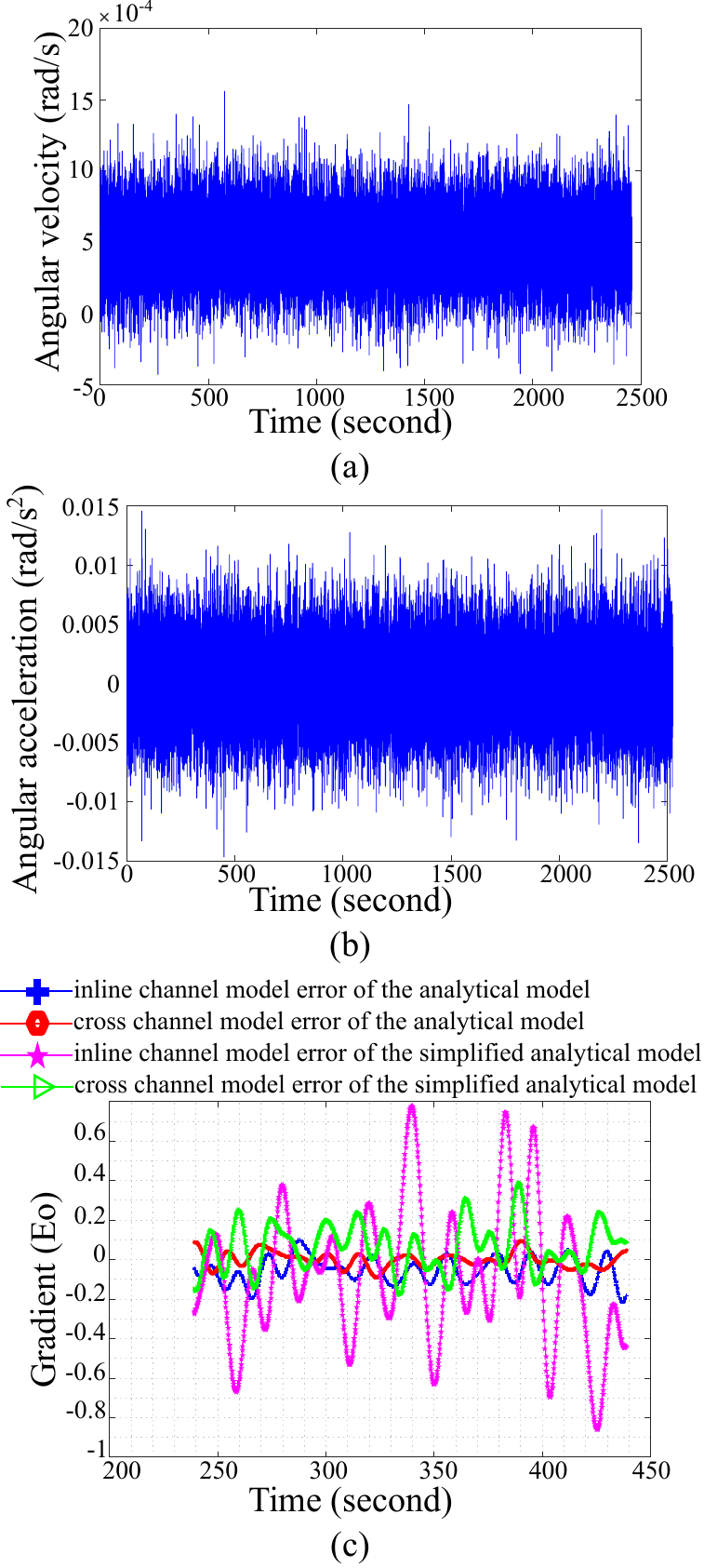}
	\caption{Results of the second turbulence simulation experiment. (a) Angular velocity the RAGG undergoing. (b) Angular acceleration the RAGG undergoing. (c) The analytical model error and the simplified analytical model error.}\label{fig10}
\end{figure}

The air turbulence acts on the aircraft, and produces vibration acceleration. As for two-engine transport aircraft, Catherines reports that RMS lateral accelerations caused by air turbulence are about 15 percent of the vertical acceleration, and about 90 percent of the vibratory energy is in the 0 $\sim$ 3~Hz frequency range during flight \cite{Catherines1975}. As the vertical acceleration caused by the air turbulence directly reflects the intensity of the air turbulence, usually the word "turbulence" is equal to the vertical acceleration. Dransfield reports that the turbulence in Cessna C208B aircraft is usually in the range of 40 $\sim$ 100~mg with a mean of about 70~mg \cite{dransfield2013performance}. In this paper, we approximate the turbulence with Gaussian noise;
let $\mu$~mg and $\sigma$~mg denote the mean and the standard deviation of the turbulence; then the turbulence is in the range of $[\mu-3\sigma,\mu+3\sigma]$~mg. To approximate the real situation, in turbulence experiments, besides the turbulence, the RAGG also suffered random angular motion. The random angular velocities of the RAGG are produced with Gaussian noise; the angular accelerations are the derivative of the angular velocities.

In the three turbulence experiments, the RMS horizontal accelerations are 15 percent of the vertical acceleration, but the RMS angular velocity components are of the same order. The turbulence and random angular velocity parameters are listed in table.\ref{tab7}; each row of the table.\ref{tab7} represents a group parameters in a turbulence experiment. The configuration parameters of the RAGG are the same as those of the dynamic experiment (listed in Table~\ref{tab5}).

The outputs of the numerical model are treated as the outputs of the actual RAGG, and the differences between the analytical model and the numerical model
is treated as the model error or model noise.
In the first turbulence experiment, the RAGG only suffered turbulence without random angular motions, the mean and the standard deviation of the turbulence are respectively, 100~mg and 10~mg; Fig.\ref{fig9}(a) shows the turbulence; Fig.\ref{fig9}(b) shows a histogram of the turbulence.
Fig.\ref{fig9}(c) shows the analytical model error and the simplified analytical model error in the first turbulence experiment. In the second turbulence experiment, the RAGG suffered both turbulence and random angular motion. The turbulence parameters in the second experiment are the same as those in the first experiment; the mean and the standard deviation of the random angular motion are respectively, 100~deg/h and 50~deg/h; Fig.\ref{fig10}(a) shows the random angular velocity; the angular accelerations are the derivative of the angular velocities; Fig.\ref{fig10}(b) shows the random angular acceleration. Fig.\ref{fig10}(c) shows the analytical model error and the simplified analytical model error in the second turbulence experiment.

In the third turbulence experiment, the random angular motion parameters are the same as those in the second experiment, but the mean of the turbulence are increased from 100~mg to 200~mg. Statistical results of the analytic model error and the simplified analytic model error in the three turbulence experiments are listed in table.\ref{tab8}; where $M_{inj}$, $\sigma_{inj}$ respectively denote the mean and standard deviation of the inline channel of the analytical model error in the j-th turbulence experiment; $M_{csj}$, $\sigma_{csj}$ respectively denote the mean and standard deviation of the cross channel of the analytical model error in the j-th turbulence experiment; $M^s_{inj}$, $\sigma^s_{inj}$ respectively denote the mean and standard deviation of the inline channel of the simplified analytical model error in the j-th turbulence experiment; $M^s_{csj}$, $\sigma^s_{csj}$ respectively denote the mean and standard deviation of the cross channel of the simplified analytical model error in the j-th turbulence experiment.

From the first turbulence experiment to the third turbulence experiment, although the turbulence and angular motion has increased, in the table.\ref{tab8}, the standard deviation of the analytic model error $\sigma_{inj}$, $\sigma_{csj}$ have no significant changes; $\sigma_{inj}$, $\sigma_{csj}$ are about 0.06~Eo and 0.03~Eo. The bandwidth of the RAGG is 0.2~Hz, the corresponding noise density of the analytic model is about 0.13 Eo/$\surd$Hz.
 In the three turbulence experiments, the standard deviation of the simplified analytic model error $\sigma^s_{inj}$, $\sigma^s_{csj}$ increase as the turbulence and angular motion increase.
This experiment result is consistent with the simplified analytic model properties; the simplified analytic model has neglected the coupling terms concerning accelerometer second-order error coefficients, angular motions, and linear motions, the coupling terms of accelerometer second-order error coefficients and angular motions; thus, the larger the turbulence and the angular motion, the greater the simplified analytic model error.
The standard deviation of the simplified analytic model error in the first experiment is $\sigma^s_{in1}=0.11~Eo$, $\sigma^s_{cs1}=0.052~Eo$, the corresponding noise density is 0.25 Eo/$\surd$Hz. In the second experiment, $\sigma^s_{in2}=0.294~Eo$, $\sigma^s_{cs2}=0.149~Eo$, the corresponding noise density is 0.66 Eo/$\surd$Hz. In the third experiment, $\sigma^s_{in3}=0.556~Eo$, $\sigma^s_{cs3}=0.281~Eo$, the corresponding noise density is 1.24 Eo/$\surd$Hz.
In turbulence experiments, although the turbulence is in the range of 100 $\sim$ 200~mg, the noise density of the simplified analytic model and the analytic model are much smaller than 7 Eo/$\surd$Hz, which suggests that using the error compensation techniques based
on the simplified analytical model and analytical model, the turbulence threshold of survey flying can be widened from current 100~mg to 200~mg.
\begin{table}[]
\centering
\caption{ The Turbulence and Random Angular Velocity Parameters in Turbulence Simulation Experiments.}
\label{tab7}
\begin{tabular}{cccc} \hline \hline
               \multicolumn{2}{c}{Linear acceleration (mg) }  & \multicolumn{2}{c}{Angular velocity (deg/h)} \\ \hline
               Mean      & Standard deviation      &  Mean         & Standard deviation          \\ \hline
                 100     & 10                      & 0             & 0                           \\
                 100     & 10                      & 100           & 50                           \\
                 200     & 10                      & 100           & 50                           \\  \hline \hline
\end{tabular}
\end{table}
\begin{table}[htbp]
  \centering
  \caption{Statistical Results of Model Error in The Three Turbulence Simulation Experiments.}
  \label{tab8}
  \setlength{\tabcolsep}{1.2mm}{
    \begin{tabular}{cccc} \hline \hline
    \multicolumn{2}{c}{Analytic model} & \multicolumn{2}{c}{Simplified analytic model}  \\ \hline
    Inline channel & Cross  channel & Inline  channel & Cross channel \\ \hline
       $M_{in1}$=-0.037~Eo     & $M_{cs1}$=-0.0004~Eo      & $M^s_{in1}$=-0.080~Eo     &  $M^s_{cs1}$=0.033~Eo     \\
       $\sigma_{in1}$=0.062~Eo & $\sigma_{cs1}$=0.031~Eo  & $\sigma^s_{in1}$=0.11~Eo &  $\sigma^s_{cs1}$=0.052~Eo  \\
       $M_{in2}$=-0.032~Eo       & $M_{cs2}$=-0.001~Eo      & $M^s_{in2}$=-0.09~Eo     &  $M^s_{cs2}$=0.015~Eo     \\
       $\sigma_{in2}$=0.068~Eo & $\sigma_{cs2}$=0.032~Eo  & $\sigma^s_{in2}$=0.294~Eo &  $\sigma^s_{cs2}$=0.149~Eo  \\
       $M_{in3}$= -0.145~Eo       & $M_{cs3}$=-0.001~Eo      & $M^s_{in3}$=-0.253~Eo     &  $M^s_{cs3}$=0.134~Eo     \\
       $\sigma_{in3}$=0.064~Eo & $\sigma_{cs3}$=0.029~Eo  & $\sigma^s_{in3}$=0.556~Eo &  $\sigma^s_{cs3}$=0.281~Eo  \\ \hline \hline
    \end{tabular}}
  \label{tab:addlabel}%
\end{table}%

\section{Conclusion}

Based on the measurement principle and configuration of the RAGG, we considered the factors of circuit gain mismatch, installation error, accelerometer scale-factor imbalance, and accelerometer second-order error coefficients. We then developed a high-precision numerical model and an analytical model of the RAGG. Based on the dynamic environment of airborne gravity gradiometry, we analyzed the magnitude of each error term, neglected those that had little impact on the RAGG sensitivity, and thereby obtained a simplified analytical model that is more suitable for RAGG error compensation. Moreover, the analytical model directly gives the error propagation mechanism of the motion of the RAGG and is helpful for developing techniques such as on-line error compensation, post-mission compensation, and fault diagnosis. Meanwhile, the numerical model will provide a tool for verifying different techniques in developing RAGG.
In the turbulence experiments, the turbulence is in the range of 100 $\sim$ 200~mg; the angular velocity is in the order of 10$^{-4}$ $\sim$ 10$^{-3}$ rad/s; the angular acceleration is in the order of 10$^{-3}$ $\sim$ 10$^{-2}$ rad/$s^2$; but the noise density of the analytic model is bout 0.13 Eo/$\surd$Hz, and that of the simplified analytic model is in the range of 0.25 $\sim$ 1.24 Eo/$\surd$Hz. In practice, after the simplified analytic model is calibrated, by recording the angular velocities, angular accelerations and accelerations of the RAGG in airborne gravity gradiometry, we can accurately compensate the measurement errors caused by turbulence and angular motion; and the turbulence threshold of survey flying may be widened from current 100~mg to 200~mg.

 \begin{acknowledgments}
This work is supported by National Key R\&D Program of China under Grant No. 2017YFC0601601, 2016YFC0303006
 and International Special Projects for Scientific and Technological Cooperation under Grant No. 2014DFR80750.
\end{acknowledgments}
\bibliography{ref}

\begin{thebibliography}{26}%
\makeatletter
\providecommand \@ifxundefined [1]{%
 \@ifx{#1\undefined}
}%
\providecommand \@ifnum [1]{%
 \ifnum #1\expandafter \@firstoftwo
 \else \expandafter \@secondoftwo
 \fi
}%
\providecommand \@ifx [1]{%
 \ifx #1\expandafter \@firstoftwo
 \else \expandafter \@secondoftwo
 \fi
}%
\providecommand \natexlab [1]{#1}%
\providecommand \enquote  [1]{``#1''}%
\providecommand \bibnamefont  [1]{#1}%
\providecommand \bibfnamefont [1]{#1}%
\providecommand \citenamefont [1]{#1}%
\providecommand \href@noop [0]{\@secondoftwo}%
\providecommand \href [0]{\begingroup \@sanitize@url \@href}%
\providecommand \@href[1]{\@@startlink{#1}\@@href}%
\providecommand \@@href[1]{\endgroup#1\@@endlink}%
\providecommand \@sanitize@url [0]{\catcode `\\12\catcode `\$12\catcode
  `\&12\catcode `\#12\catcode `\^12\catcode `\_12\catcode `\%12\relax}%
\providecommand \@@startlink[1]{}%
\providecommand \@@endlink[0]{}%
\providecommand \url  [0]{\begingroup\@sanitize@url \@url }%
\providecommand \@url [1]{\endgroup\@href {#1}{\urlprefix }}%
\providecommand \urlprefix  [0]{URL }%
\providecommand \Eprint [0]{\href }%
\providecommand \doibase [0]{http://dx.doi.org/}%
\providecommand \selectlanguage [0]{\@gobble}%
\providecommand \bibinfo  [0]{\@secondoftwo}%
\providecommand \bibfield  [0]{\@secondoftwo}%
\providecommand \translation [1]{[#1]}%
\providecommand \BibitemOpen [0]{}%
\providecommand \bibitemStop [0]{}%
\providecommand \bibitemNoStop [0]{.\EOS\space}%
\providecommand \EOS [0]{\spacefactor3000\relax}%
\providecommand \BibitemShut  [1]{\csname bibitem#1\endcsname}%
\let\auto@bib@innerbib\@empty
\bibitem [{\citenamefont {Tang}\ \emph {et~al.}(2018)\citenamefont {Tang},
  \citenamefont {Hu}, \citenamefont {Ren},\ and\ \citenamefont
  {Chen}}]{Tang2018}%
  \BibitemOpen
  \bibfield  {author} {\bibinfo {author} {\bibfnamefont {J.}~\bibnamefont
  {Tang}}, \bibinfo {author} {\bibfnamefont {S.}~\bibnamefont {Hu}}, \bibinfo
  {author} {\bibfnamefont {Z.}~\bibnamefont {Ren}}, \ and\ \bibinfo {author}
  {\bibfnamefont {C.}~\bibnamefont {Chen}},\ }\href {\doibase
  10.1109/LGRS.2017.2784837} {\bibfield  {journal} {\bibinfo  {journal} {IEEE
  Geoscience and Remote Sensing Letters}\ }\textbf {\bibinfo {volume} {15}},\
  \bibinfo {pages} {247} (\bibinfo {year} {2018})}\BibitemShut {NoStop}%
\bibitem [{\citenamefont {Yan}, \citenamefont {Ma},\ and\ \citenamefont
  {Tian}(2015)}]{Yan2015}%
  \BibitemOpen
  \bibfield  {author} {\bibinfo {author} {\bibfnamefont {Z.}~\bibnamefont
  {Yan}}, \bibinfo {author} {\bibfnamefont {J.}~\bibnamefont {Ma}}, \ and\
  \bibinfo {author} {\bibfnamefont {J.}~\bibnamefont {Tian}},\ }\href {\doibase
  10.1109/LGRS.2015.2388772} {\bibfield  {journal} {\bibinfo  {journal} {IEEE
  Geoscience and Remote Sensing Letters}\ }\textbf {\bibinfo {volume} {12}},\
  \bibinfo {pages} {1214} (\bibinfo {year} {2015})}\BibitemShut {NoStop}%
\bibitem [{\citenamefont {Kahn}\ and\ \citenamefont
  {Bun}(1985)}]{Kahn1985IToGaRS}%
  \BibitemOpen
  \bibfield  {author} {\bibinfo {author} {\bibfnamefont {W.~D.}\ \bibnamefont
  {Kahn}}\ and\ \bibinfo {author} {\bibfnamefont {F.~O.~V.}\ \bibnamefont
  {Bun}},\ }\href {\doibase 10.1109/TGRS.1985.289445} {\bibfield  {journal}
  {\bibinfo  {journal} {IEEE Transactions on Geoscience and Remote Sensing}\
  }\textbf {\bibinfo {volume} {GE-23}},\ \bibinfo {pages} {527} (\bibinfo
  {year} {1985})}\BibitemShut {NoStop}%
\bibitem [{\citenamefont {Welker}, \citenamefont {Pachter},\ and\ \citenamefont
  {Huffman}(2013)}]{welker2013gravity}%
  \BibitemOpen
  \bibfield  {author} {\bibinfo {author} {\bibfnamefont {T.~C.}\ \bibnamefont
  {Welker}}, \bibinfo {author} {\bibfnamefont {M.}~\bibnamefont {Pachter}}, \
  and\ \bibinfo {author} {\bibfnamefont {R.~E.}\ \bibnamefont {Huffman}},\ }in\
  \href {\doibase 10.23919/ECC.2013.6669109} {\emph {\bibinfo {booktitle}
  {Control Conference (ECC), 2013 European}}}\ (\bibinfo {organization}
  {IEEE},\ \bibinfo {year} {2013})\ pp.\ \bibinfo {pages}
  {846--851}\BibitemShut {NoStop}%
\bibitem [{\citenamefont {Araya}\ \emph {et~al.}(2012)\citenamefont {Araya},
  \citenamefont {Kanazawa}, \citenamefont {Shinohara}, \citenamefont {Yamada},
  \citenamefont {Fujimoto}, \citenamefont {Iizasa},\ and\ \citenamefont
  {Ishihara}}]{araya2012gravity}%
  \BibitemOpen
  \bibfield  {author} {\bibinfo {author} {\bibfnamefont {A.}~\bibnamefont
  {Araya}}, \bibinfo {author} {\bibfnamefont {T.}~\bibnamefont {Kanazawa}},
  \bibinfo {author} {\bibfnamefont {M.}~\bibnamefont {Shinohara}}, \bibinfo
  {author} {\bibfnamefont {T.}~\bibnamefont {Yamada}}, \bibinfo {author}
  {\bibfnamefont {H.}~\bibnamefont {Fujimoto}}, \bibinfo {author}
  {\bibfnamefont {K.}~\bibnamefont {Iizasa}}, \ and\ \bibinfo {author}
  {\bibfnamefont {T.}~\bibnamefont {Ishihara}},\ }in\ \href {\doibase
  10.1109/OCEANS.2012.6405114} {\emph {\bibinfo {booktitle} {Oceans, 2012}}}\
  (\bibinfo {organization} {IEEE},\ \bibinfo {year} {2012})\ pp.\ \bibinfo
  {pages} {1--4}\BibitemShut {NoStop}%
\bibitem [{\citenamefont {Jekeli}(2006{\natexlab{a}})}]{Jekeli2006Precision}%
  \BibitemOpen
  \bibfield  {author} {\bibinfo {author} {\bibfnamefont {C.}~\bibnamefont
  {Jekeli}},\ }\href {\doibase 10.2514/1.15368} {\bibfield  {journal} {\bibinfo
   {journal} {Journal of Guidance Control \& Dynamics}\ }\textbf {\bibinfo
  {volume} {29}},\ \bibinfo {pages} {704} (\bibinfo {year}
  {2006}{\natexlab{a}})}\BibitemShut {NoStop}%
\bibitem [{\citenamefont {Rogers}(2009)}]{rogers2009investigation}%
  \BibitemOpen
  \bibfield  {author} {\bibinfo {author} {\bibfnamefont {M.~M.}\ \bibnamefont
  {Rogers}},\ }\href {https://archive.org/details/DTIC_ADA496707} {\enquote
  {\bibinfo {title} {An investigation into the feasibility of using a modern
  gravity gradient instrument for passive aircraft navigation and terrain
  avoidance},}\ }\bibinfo {type} {techreport}\ \bibinfo {number} {ADA496707}\
  (\bibinfo  {institution} {Air Force Institute of Technology, Wright-Patterson
  Air Force Base, Ohio, Graduate School of Engineering and Management},\
  \bibinfo {year} {2009})\BibitemShut {NoStop}%
\bibitem [{\citenamefont {Annecchione}\ \emph {et~al.}(2007)\citenamefont
  {Annecchione}, \citenamefont {Moody}, \citenamefont {Carroll}, \citenamefont
  {Dickson},\ and\ \citenamefont {Main}}]{annecchione2007benefits}%
  \BibitemOpen
  \bibfield  {author} {\bibinfo {author} {\bibfnamefont {M.}~\bibnamefont
  {Annecchione}}, \bibinfo {author} {\bibfnamefont {M.}~\bibnamefont {Moody}},
  \bibinfo {author} {\bibfnamefont {K.}~\bibnamefont {Carroll}}, \bibinfo
  {author} {\bibfnamefont {D.}~\bibnamefont {Dickson}}, \ and\ \bibinfo
  {author} {\bibfnamefont {B.}~\bibnamefont {Main}},\ }in\ \href
  {https://www.911metallurgist.com/blog/wp-content/uploads/2015/10/Benefits-of-a-High-Performance-Airborne-Gravity-Gradiometer-for-Resource-Exploration.pdf}
  {\emph {\bibinfo {booktitle} {Fifth Decennial International Conference on
  Mineral Exploration}}}\ (\bibinfo {year} {2007})\ pp.\ \bibinfo {pages}
  {889--893}\BibitemShut {NoStop}%
\bibitem [{\citenamefont {Affleck}\ and\ \citenamefont
  {Jircitano}(1990)}]{66158}%
  \BibitemOpen
  \bibfield  {author} {\bibinfo {author} {\bibfnamefont {C.~A.}\ \bibnamefont
  {Affleck}}\ and\ \bibinfo {author} {\bibfnamefont {A.}~\bibnamefont
  {Jircitano}},\ }in\ \href {\doibase 10.1109/PLANS.1990.66158} {\emph
  {\bibinfo {booktitle} {IEEE Symposium on Position Location and Navigation. A
  Decade of Excellence in the Navigation Sciences}}}\ (\bibinfo {year} {1990})\
  pp.\ \bibinfo {pages} {60--66}\BibitemShut {NoStop}%
\bibitem [{\citenamefont {Paik}(2007)}]{Paik2007Geodesy}%
  \BibitemOpen
  \bibfield  {author} {\bibinfo {author} {\bibfnamefont {H.~J.}\ \bibnamefont
  {Paik}},\ }\href {\doibase 10.1109/TGRS.1985.289444} {\bibfield  {journal}
  {\bibinfo  {journal} {IEEE Transactions on Geoscience \& Remote Sensing}\
  }\textbf {\bibinfo {volume} {GE-23}},\ \bibinfo {pages} {524} (\bibinfo
  {year} {2007})}\BibitemShut {NoStop}%
\bibitem [{\citenamefont {Anstie}\ \emph {et~al.}(2009)\citenamefont {Anstie},
  \citenamefont {Aravanis}, \citenamefont {Haederle}, \citenamefont {Mann},
  \citenamefont {McIntosh}, \citenamefont {Smith}, \citenamefont {Van~Kann},
  \citenamefont {Wells},\ and\ \citenamefont {Winterflood}}]{anstie2009vk}%
  \BibitemOpen
  \bibfield  {author} {\bibinfo {author} {\bibfnamefont {J.}~\bibnamefont
  {Anstie}}, \bibinfo {author} {\bibfnamefont {T.}~\bibnamefont {Aravanis}},
  \bibinfo {author} {\bibfnamefont {M.}~\bibnamefont {Haederle}}, \bibinfo
  {author} {\bibfnamefont {A.}~\bibnamefont {Mann}}, \bibinfo {author}
  {\bibfnamefont {S.}~\bibnamefont {McIntosh}}, \bibinfo {author}
  {\bibfnamefont {R.}~\bibnamefont {Smith}}, \bibinfo {author} {\bibfnamefont
  {F.}~\bibnamefont {Van~Kann}}, \bibinfo {author} {\bibfnamefont
  {G.}~\bibnamefont {Wells}}, \ and\ \bibinfo {author} {\bibfnamefont
  {J.}~\bibnamefont {Winterflood}},\ }\href {\doibase 10.1071/ASEG2009ab084}
  {\bibfield  {journal} {\bibinfo  {journal} {ASEG Extended Abstracts}\
  }\textbf {\bibinfo {volume} {2009}},\ \bibinfo {pages} {1} (\bibinfo {year}
  {2009})}\BibitemShut {NoStop}%
\bibitem [{\citenamefont {Difrancesco}(2007)}]{difrancesco2007advances}%
  \BibitemOpen
  \bibfield  {author} {\bibinfo {author} {\bibfnamefont {D.}~\bibnamefont
  {Difrancesco}},\ }in\ \href {\doibase 10.1071/ASEG2007ab034} {\emph {\bibinfo
  {booktitle} {EGM 2007 international workshop}}},\ Vol.\ \bibinfo {volume}
  {2007}\ (\bibinfo  {publisher} {{CSIRO} Publishing},\ \bibinfo {year}
  {2007})\ p.~\bibinfo {pages} {1}\BibitemShut {NoStop}%
\bibitem [{\citenamefont {Hao}, \citenamefont {Tijing},\ and\ \citenamefont
  {Tao}(2013)}]{hao2013design}%
  \BibitemOpen
  \bibfield  {author} {\bibinfo {author} {\bibfnamefont {D.}~\bibnamefont
  {Hao}}, \bibinfo {author} {\bibfnamefont {C.}~\bibnamefont {Tijing}}, \ and\
  \bibinfo {author} {\bibfnamefont {S.}~\bibnamefont {Tao}},\ }in\ \href
  {\doibase 10.1109/ICEMI.2013.6743105} {\emph {\bibinfo {booktitle}
  {Electronic Measurement \& Instruments (ICEMI), 2013 IEEE 11th International
  Conference on}}},\ Vol.~\bibinfo {volume} {1}\ (\bibinfo {organization}
  {IEEE},\ \bibinfo {year} {2013})\ pp.\ \bibinfo {pages}
  {396--398}\BibitemShut {NoStop}%
\bibitem [{\citenamefont {Moody}, \citenamefont {Chan},\ and\ \citenamefont
  {Paik}(1983)}]{moody1983preliminary}%
  \BibitemOpen
  \bibfield  {author} {\bibinfo {author} {\bibfnamefont {M.}~\bibnamefont
  {Moody}}, \bibinfo {author} {\bibfnamefont {H.}~\bibnamefont {Chan}}, \ and\
  \bibinfo {author} {\bibfnamefont {H.}~\bibnamefont {Paik}},\ }\href {\doibase
  10.1109/tmag.1983.1062415} {\bibfield  {journal} {\bibinfo  {journal} {IEEE
  Transactions on Magnetics}\ }\textbf {\bibinfo {volume} {19}},\ \bibinfo
  {pages} {461} (\bibinfo {year} {1983})}\BibitemShut {NoStop}%
\bibitem [{\citenamefont {Moody}(2011)}]{moody2011superconducting}%
  \BibitemOpen
  \bibfield  {author} {\bibinfo {author} {\bibfnamefont {M.}~\bibnamefont
  {Moody}},\ }\href {\doibase 10.1063/1.3632114} {\bibfield  {journal}
  {\bibinfo  {journal} {Review of Scientific Instruments}\ }\textbf {\bibinfo
  {volume} {82}},\ \bibinfo {pages} {094501} (\bibinfo {year}
  {2011})}\BibitemShut {NoStop}%
\bibitem [{\citenamefont {Liu}, \citenamefont {Pike},\ and\ \citenamefont
  {Dou}(2014)}]{liu2014design}%
  \BibitemOpen
  \bibfield  {author} {\bibinfo {author} {\bibfnamefont {H.}~\bibnamefont
  {Liu}}, \bibinfo {author} {\bibfnamefont {W.}~\bibnamefont {Pike}}, \ and\
  \bibinfo {author} {\bibfnamefont {G.}~\bibnamefont {Dou}},\ }in\ \href
  {\doibase 10.1109/ICSENS.2014.6985327} {\emph {\bibinfo {booktitle} {SENSORS,
  2014 IEEE}}}\ (\bibinfo {organization} {IEEE},\ \bibinfo {year} {2014})\ pp.\
  \bibinfo {pages} {1611--1614}\BibitemShut {NoStop}%
\bibitem [{\citenamefont {Lee}(2001)}]{lee2001falcon}%
  \BibitemOpen
  \bibfield  {author} {\bibinfo {author} {\bibfnamefont {J.~B.}\ \bibnamefont
  {Lee}},\ }\href {\doibase 10.1071/aseg2001ab068} {\bibfield  {journal}
  {\bibinfo  {journal} {Exploration Geophysics}\ }\textbf {\bibinfo {volume}
  {32}},\ \bibinfo {pages} {247} (\bibinfo {year} {2001})}\BibitemShut
  {NoStop}%
\bibitem [{\citenamefont {Li}\ and\ \citenamefont {Cai}(2010)}]{HaibingLi2010}%
  \BibitemOpen
  \bibfield  {author} {\bibinfo {author} {\bibfnamefont {H.}~\bibnamefont
  {Li}}\ and\ \bibinfo {author} {\bibfnamefont {T.}~\bibnamefont {Cai}},\
  }\href {\doibase 10.3969/j.issn.1001-0505.2010.03.016} {\bibfield  {journal}
  {\bibinfo  {journal} {Journal of Southeast University (Natural Science
  Edition)}\ }\textbf {\bibinfo {volume} {40}},\ \bibinfo {pages} {517}
  (\bibinfo {year} {2010})}\BibitemShut {NoStop}%
\bibitem [{\citenamefont {Yu}\ and\ \citenamefont
  {Cai}(2018)}]{yu2018calibration}%
  \BibitemOpen
  \bibfield  {author} {\bibinfo {author} {\bibfnamefont {M.}~\bibnamefont
  {Yu}}\ and\ \bibinfo {author} {\bibfnamefont {T.}~\bibnamefont {Cai}},\
  }\href {\doibase 10.1063/1.5018839} {\bibfield  {journal} {\bibinfo
  {journal} {Review of Scientific Instruments}\ }\textbf {\bibinfo {volume}
  {89}},\ \bibinfo {pages} {054502} (\bibinfo {year} {2018})}\BibitemShut
  {NoStop}%
\bibitem [{\citenamefont {Metzger}(1977)}]{metzger1977recent}%
  \BibitemOpen
  \bibfield  {author} {\bibinfo {author} {\bibfnamefont {E.}~\bibnamefont
  {Metzger}},\ }in\ \href {\doibase 10.2514/6.1977-1081} {\emph {\bibinfo
  {booktitle} {Guidance and Control Conference}}}\ (\bibinfo {year} {1977})\
  p.\ \bibinfo {pages} {1081}\BibitemShut {NoStop}%
\bibitem [{\citenamefont {Heard}(1988)}]{heard1988gravity}%
  \BibitemOpen
  \bibfield  {author} {\bibinfo {author} {\bibfnamefont {H.}~\bibnamefont
  {Heard}},\ }\href {\doibase 10.1029/88EO00070} {\bibfield  {journal}
  {\bibinfo  {journal} {Eos}\ }\textbf {\bibinfo {volume} {69}} (\bibinfo
  {year} {1988}),\ 10.1029/88EO00070}\BibitemShut {NoStop}%
\bibitem [{\citenamefont {Geospace}(2004)}]{geospace2004final}%
  \BibitemOpen
  \bibfield  {author} {\bibinfo {author} {\bibfnamefont {B.}~\bibnamefont
  {Geospace}},\ }\href@noop {} {\bibfield  {journal} {\bibinfo  {journal} {Rice
  University, Houston, Texas}\ } (\bibinfo {year} {2004})}\BibitemShut
  {NoStop}%
\bibitem [{\citenamefont {Dransfield}\ and\ \citenamefont
  {Christensen}(2013)}]{dransfield2013performance}%
  \BibitemOpen
  \bibfield  {author} {\bibinfo {author} {\bibfnamefont {M.~H.}\ \bibnamefont
  {Dransfield}}\ and\ \bibinfo {author} {\bibfnamefont {A.~N.}\ \bibnamefont
  {Christensen}},\ }\href {\doibase 10.1190/tle32080908.1} {\bibfield
  {journal} {\bibinfo  {journal} {The Leading Edge}\ }\textbf {\bibinfo
  {volume} {32}},\ \bibinfo {pages} {908} (\bibinfo {year} {2013})}\BibitemShut
  {NoStop}%
\bibitem [{\citenamefont {Jekeli}(2006{\natexlab{b}})}]{jekeli2006airborne}%
  \BibitemOpen
  \bibfield  {author} {\bibinfo {author} {\bibfnamefont {C.}~\bibnamefont
  {Jekeli}},\ }\href {\doibase 10.1007/s10712-005-3826-4} {\bibfield  {journal}
  {\bibinfo  {journal} {Surveys in Geophysics}\ }\textbf {\bibinfo {volume}
  {27}},\ \bibinfo {pages} {257} (\bibinfo {year}
  {2006}{\natexlab{b}})}\BibitemShut {NoStop}%
\bibitem [{\citenamefont {Ma}(2012)}]{Ma2012}%
  \BibitemOpen
  \bibfield  {author} {\bibinfo {author} {\bibfnamefont {C.}~\bibnamefont
  {Ma}},\ }\emph {\bibinfo {title} {Error analysis of rotating accelerometer
  gravity gradiometer}},\ \href
  {http://www.wanfangdata.com.cn/details/detail.do?_type=degree&id=Y2247809}
  {Master's thesis},\ \bibinfo  {school} {Southeast University} (\bibinfo
  {year} {2012})\BibitemShut {NoStop}%
\bibitem [{\citenamefont {Catherines}, \citenamefont {Mixson},\ and\
  \citenamefont {Scholl}(1975)}]{Catherines1975}%
  \BibitemOpen
  \bibfield  {author} {\bibinfo {author} {\bibfnamefont {J.~J.}\ \bibnamefont
  {Catherines}}, \bibinfo {author} {\bibfnamefont {J.~S.}\ \bibnamefont
  {Mixson}}, \ and\ \bibinfo {author} {\bibfnamefont {H.~F.}\ \bibnamefont
  {Scholl}},\ }\href {https://ntrs.nasa.gov/search.jsp?R=19750020964} {\enquote
  {\bibinfo {title} {Vibrations measured in the passenger cabins of two jet
  transport aircraft},}\ }\bibinfo {type} {techreport}\ \bibinfo {number}
  {NASA-TN-D-7923, L-9531}\ (\bibinfo  {institution} {NASA Langley Research
  Center, Hampton, VA, United States},\ \bibinfo {address} {Washington, United
  States},\ \bibinfo {year} {1975})\BibitemShut {NoStop}%
\end{thebibliography}%

\end{document}